\documentclass[12pt]{article}
\usepackage[a4paper, left=1.0in, right=1.0in, top=1in, bottom=1in]{geometry}
\usepackage{latexsym}
\usepackage{amssymb}
\usepackage{amsmath}
\usepackage[T1]{fontenc}
\usepackage[american]{babel}
\usepackage[affil-it]{authblk}
\usepackage{url}
\usepackage{amsthm}
\usepackage[round]{natbib}
\usepackage{placeins}
\usepackage{color}
\usepackage{lipsum}
\usepackage{graphicx}
\usepackage{epstopdf}
\usepackage{multirow}
\usepackage{tablefootnote}
\usepackage{setspace}
\usepackage{pbox}
\usepackage{tabularx,ragged2e,booktabs}
\usepackage{calc}
\usepackage{verbatim}
\usepackage{pdflscape}
\usepackage{arydshln}
\usepackage{array}
\usepackage[overload]{empheq}
\usepackage{tikz}
\usepackage[font=footnotesize, hypcap=false]{caption}
\usepackage{pgfplots}
\pgfplotsset{compat=1.18}
\usepackage{subfigure}
\usepackage{framed}
\usepackage{pdflscape}
\usepackage{soul}
\usepackage{changepage}
\usepackage[toc,title]{appendix}
\usetikzlibrary{calc,decorations.pathreplacing}
\newcolumntype{s}{>{\hsize=.5\hsize}X}
\newcommand\blfootnote[1]{%
	\begingroup
	\renewcommand\thefootnote{}\footnote{#1}%
	\addtocounter{footnote}{-1}%
	\endgroup
}
\usepackage[hidelinks]{hyperref}
\usepackage{float}
\usepackage{tablefootnote}
\usepackage{enumitem}
\newcommand{\heatmapcell}[1]{%
  \pgfmathsetmacro{\value}{#1}%
  \pgfmathsetmacro{\greenval}{min(1, max(0, (\value - 0.5)/0.5))}%
  \pgfmathsetmacro{\fadeval}{1 - \greenval}%
  \xdef\colorname{heatmapcolor\value}%
  \xdef\colorvalue{\fadeval,1,\fadeval}%
  \expandafter\definecolor\csname \colorname\endcsname{rgb}{\colorvalue}%
  \cellcolor{\colorname}#1%
}

\begin{document}
\setlength{\marginparwidth}{5pc}
\title{ Central Bank Digital Currency, Flight-to-Quality, and Bank-Runs in an Agent-Based Model \\ }
    \author[a]{Emilio Barucci}
    \affil[a]{\footnotesize Department of  Mathematics, Politecnico di Milano, \texttt{emilio.barucci@polimi.it} }
    \author[b,c]{Andrea Gurgone\footnote{The support from the UKRI Grant entitled "The Large Agent Collider: Robust agent-based modelling as scale" awarded to Prof. Wooldridge (reference EP/W002949/1) is gratefully acknowledged. In addition, the authors would like to acknowledge the use of the University of Oxford Advanced Research Computing (ARC) facility in carrying out this work. \url{http://dx.doi.org/10.5281/zenodo.22558}}}
    \affil[b]{\footnotesize Department of Computer Science, University of Oxford, \texttt{andrea.gurgone@cs.ox.ac.uk}}
    \affil[c]{\footnotesize Institute for New Economic Thinking, University of Oxford}
    \author[e,f]{Giulia Iori}
    \affil[e]{\footnotesize Department of  Economics, Ca' Foscari University of Venice, \texttt{giulia.iori@unive.it} }
    \affil[f]{\footnotesize Department of  Economics, City St. George's, University of London}    \author[a]{Michele Azzone}

    \normalsize
    \date{}   
    \maketitle
    \blfootnote{\texttt{Corresponding author andrea.gurgone@cs.ox.ac.uk}
}
    \vspace{-3em}
    \begin{abstract}
    We analyse financial stability and welfare impacts associated with the introduction of a Central Bank Digital Currency (CBDC) in a macroeconomic agent-based model. The model considers firms, banks, and households interacting on labour, goods, credit, and interbank markets. Households move their liquidity from deposits to CBDC based on the perceived riskiness of their banks. 
    We find that the introduction of CBDC exacerbates bank-runs and may lead to financial instability phenomena. The effect can be changed by introducing a limit on CBDC holdings. The adoption of CBDC has little effect on macroeconomic variables but the interest rate on loans to firms goes up and credit goes down in a limited way. CBDC leads to a redistribution of wealth from firms and banks to households with a higher bank default rate. 
    CBDC may have negative welfare effects, but a bound on holding enables a welfare improvement.
    
    
    
    \end{abstract}
    \clearpage
    
    \section{Introduction}

    The debate on issuing retail Central Bank Digital Currency (CBDC) focuses primarily on welfare and financial stability implications.
    The literature has highlighted the main trade-off. On one hand, CBDC aligns with citizens' preferences for digital payments with more efficient exchanges; on the other hand, there are concerns regarding the substitution of deposits with digital money, which may lead to financial disintermediation, negative effects for the real economy, and instability under stress. 
    
    The existing literature on CBDC highlights that a careful design is required to find the balance along the above trade-off, see \citet{AGUR_ET_ALL,AND,ASSE,BRUN_NIEP,KEI_MON,KEI_SAN,KIM_KWON,WILL,WILL2}. In this paper, we contribute to this debate through the analysis of an agent-based model inspired by \citet{delli2011macroeconomics}, see also \citet{gurgone2022macroprudential,gurgone2018effects}. 
    We focus on flight-to-quality by households who may  substitute deposits  with CBDC. They substitute deposits with CBDC on the basis of the riskiness of their banks, fearing that they may default. Unlike the analysis provided in the above papers, the agent-based model allows us, at the same time, to endogenously determine the riskiness of banks, the bank-run of depositors switching from deposits to CBDC, and the default of banks. This feature of the model allows us to evaluate the financial stability and welfare implications associated with CBDC in a more comprehensive way than the previous literature.

Extensive household adoption of CBDC could cause financial disintermediation, resulting in negative welfare effects.
    Remuneration of CBDC and bounds to its adoption play a key role in keeping this potential problem under control.
    \citet{AGUR_ET_ALL,KEI_SAN} show that non positive remuneration can help mitigate financial disintermediation. \citet{AND} predicts a zero uptake of CBDC if its remuneration is below that of reserves held by the Central Bank (CB).
    In the Euro area, \citet{BURL} estimate that the welfare-maximizing amount of CBDC lies between 15 and 45\% of quarterly GDP in equilibrium, assuming a non positive remuneration scheme and holding bounds.
    \citet{ADAL} demonstrate that introducing a holding limit -- such as the €3,000 cap per CBDC account currently envisaged for the digital euro -- would allow aggregate CBDC holdings to reach an amount comparable to the current stock of cash (approximately one trillion euros) without generating significant disruptions to the banking system or the broader economy. Commercial banks could meet the resulting demand for CBDC by utilizing their existing reserves and by adjusting their liquidity positions through the interbank market.
    The need to impose a bound on holdings seems to be a key ingredient for a safe introduction of CBDC with a limited effect on bank profitability, see  \citet{AZZ_BAR,NYFFE}.  

    As far as financial stability is concerned, several papers point out that CBDC effectively increases the probability of a bank-run because depositors may easily switch to a safe asset under financial stress, but the welfare consequences are not so obvious. 
    \citet{WILL2} shows that a bank-run with CBDC is more likely but it will hurt households less than in the case where only physical currency is in place because households can use CBDC to perform transactions in more situations. In other words, banking panics are more frequent but are less disruptive.     
    \citet{KEI_MON} show that the increased risk of bank-runs associated with CBDC introduction can be mitigated by two mechanisms. First, banks are likely to reduce their maturity transformation, making them less vulnerable to runs. Second, by monitoring the flow of funds into CBDC, policymakers can identify and intervene in weak banks at an earlier stage.
    \citet{KIM_KWON} show that CBDC may lead to a more resilient financial system if the CB lends the deposits in the CBDC account to banks. In a bank-run model, \citet{FERN_ET_ALL} show that the CB has the capability to deter runs and may become a monopolist for deposits, endangering credit supply to the real economy.
    \citet{AHN} extend the \citet{DI_DY} model to a setting including the CBDC. The bank-run analysis is performed through a global games approach showing that the relationship between bank fragility and CBDC remuneration is U-shaped.
     
    In our model, we abstract from remuneration of CBDC, which is set to be equal to the deposit rate and we also abstract from anonymity/digitalization features and network effects of CBDC as modeled in \citet{BA_BR_MA}. We concentrate on the flight-to-quality phenomenon, with depositors transferring their liquidity to CBDC depending on the riskiness of their banks. As far as we know, this is the first paper addressing the connection between bank specific riskiness and adoption of CBDC in an endogenous way. In the existing literature, a bank-run is considered in the spirit of \citet{DI_DY}: \citet{FERN_ET_ALL,WILL} consider bank-run as a self-fulfilling prophecy; in \citet{KEI_MON,KIM_KWON,AHN} bank weakness is driven by an exogenous variable. In all of these papers, withdrawal of deposits is not related to the real economy. Instead, in our setting, a bank-run towards CBDC is motivated by the riskiness of banks which reflects conditions of the whole economy.

    \citet{ADAL} provides the most similar analysis to ours simulating the introduction of CBDC on real data for the Euro area. They assume that when the run has been triggered, citizens substitute some of their deposits with CBDC. The decision is $0-1$ and the probability of bank default, which is exogenous, increases over time. Heterogeneity is introduced assuming a normally distributed idiosyncratic component specific to each agent.  
    If demand of CBDC is unconstrained, then the scale and the speed of a system wide bank-run would increase. A hard limit on individual CBDC holdings would avoid the rise of a system wide bank-run. 

    In our setting, we investigate whether the possibility to substitute deposits for CBDC may ignite a system wide bank-run with a cascade of bank defaults and negative welfare implications.

    We consider five different rules determining the fraction of deposits converted into CBDC: flat fraction (CBDC0 rule),  fraction dependent on bank's riskiness with a loose upper-bound (CBDC1 rule) yielding almost unconstrained substitution, fraction dependent on bank's riskiness with a tight upper-bound (CBDC2 and 3 rule) and a deposit insurance scheme (CBDC4 rule).
     The first two rules represent our central scenarios: fixed fraction (an amount of CBDC similar to cash) and     conversion driven by a flight-to-quality fearing the default of banks.
    CBDC2-4 rules allow us to investigate the role played by bounds on the adoption of CBDC and deposit insurance schemes. The model is calibrated for the Euro area.

    Assuming a fixed 10\% fraction of deposits converted into CBDC (CBDC0 rule), the substitution of deposits for CBDC has limited effects on the macroeconomy (real GDP and unemployment), but the interest rate of loans of banks to firms goes up and credit to firms goes down in a limited way. 
When deposits are substituted based on banks’ risk profiles and a relatively loose upper bound is applied (CBDC1 rule), the model yields, on average, pronounced negative effects and heightened volatility.
   Introducing a bound on CBDC adoption, together with a deposit insurance scheme, effectively mitigates these effects, producing outcomes similar to those observed under the CBDC0 rule.

    CBDC leads to a redistribution of wealth from firms and banks to households, and to a higher banks' default rate. Banks cope with households' requests by exchanging liquidity among themselves in the interbank market and interbank lending goes up significantly. The increase of bank defaults is due to a stronger transmission in the interbank market (banks-banks default) and the liquidation of assets by banks.
    
    Our analysis shows that unconstrained CBDC adoption may significantly hinder the economy with a lower growth rate and more pronounced fluctuations. CBDC may ignite a digital bank-run hampering financial stability. However,  a reasonable bound on CBDC adoption (30\% of deposits) renders almost no effect in comparison to the economy without CBDC.   

    The optimal amount of CBDC intake by households is evaluated through a social welfare analysis. Varying the upper-bound on CBDC adoption, we are able to show that a 40\% bound on the fraction of deposits converted into CBDC seems to be the optimal choice. It seems that social welfare is negatively affected only in case of a massive conversion of deposits into CBDC.

    The paper is organized as follows. In Section \ref{MOD}, we present the model with five subsections dealing with households, labour market, government and central bank, business sector, banking sector. 
    In Section \ref{ADOP}, we consider different rules concerning the adoption of CBDC. In Section \ref{sec:results}, we provide and discuss the simulations. In Section \ref{sec.soc}, we develop the social welfare analysis. In Section \ref{sec:conclusion}, we draw our conclusions.

    \section{The model}
    \label{MOD}
    The model builds on the one proposed in \citet{gurgone2022macroprudential}. The main differences are related to the introduction of CBDC which affects the allocation of wealth by households and the balance sheets of banks and CB.
    
    The economy consists of five types of agents: households, firms, banks, government, and CB. Interactions occur in different markets: firms and households meet in the goods and labour markets; firms borrow from banks in the credit market; banks exchange liquidity in the interbank market. The CB buys government-issued  debt securities (bonds) in the bond market. The government’s role is limited to making transfer payments to households, funded by taxes, or issuing government bonds. The CB creates liquidity by purchasing government bonds. On the liability side it holds bank reserves and CBDC. Households earn income from wages, assets, and transfers, which they use for consumption, asset accumulation (deposits and CBDC), and to pay taxes. They are represented by a trade union in wage negotiations and own shares of firms and banks, receiving dividends as asset income. Firms borrow from banks to pay wages, hire workers, produce, and sell goods. Banks hold government bonds, provide credit under regulatory constraints and manage liquidity through the interbank market.

    \subsection{Households}
    \label{sec.households}

    The household sector is made up of $N^H$ units indexed by $i$. 	
    The net wealth ($nw$) of the $i$-th household consists of its holdings of deposits and CBDC: 

    \begin{equation}
    \label{eq.nw_H}
    nw^{H}_{i,t}=D_{i,t}^H+CBDC_{i,t}\;\;,
    \end{equation}
    The law of motion of net wealth is
    \begin{equation}\label{eq.nw_H2}
    nw^{H}_{i,t}=nw^{H}_{i,t-1}+S_{i,t}^{H}\;\;,
    \end{equation}
    where  saving $S_{i,t}^{H}$ is given by 
    \begin{align}\nonumber
    S_{i,t}^{H}&=D_{i,t-1}^H r^{D}+CBDC_{i,t-1}r^{CBDC}\\&+(1-\theta^H)\left(W_{t-1} N^H_{i,t-1}+\sum_{k=f,b}
    \delta^k_t\Pi^k_{i,t-1}\right)-C_{i,t-1}+\frac{G_t}{N^H}.
    \end{align}

    The disposable income of households at the beginning of period $t$ is given by the flow of interest on deposits held in the previous period ($D_{i,t-1}^H r^{D}$), interest on CBDC held in the previous period ($CBDC_{i,t-1}r^{CBDC}$), the income available at the end of the period $t-1$ taxed at the rate $\theta^H$  plus the government transfers to households in $t$ ($\frac{G_t}{N^H}$). 
    Income at time $t-1$ is made up of worked hours $N^H$ multiplied by the wage rate $W$ plus the dividend 
    share $\delta^k$ of net profits of firms and 
    banks $\Pi^k$, $k=f,b$, where $f$ and $b$ respectively refer to firms and banks. Saving at the beginning of period $t$ is given by disposable income minus consumption at the end of period $t-1$ ($C_{i,t-1}$).

    As far as consumption is concerned, a permanent income rule is considered: household $i$ wants to consume a fraction $c_1$ of  labour income plus government transfers and a fraction $c_2$ of wealth:
    \begin{equation}\label{eq.C}
    C_{i,t}^d=c_{1}\left[(1-\theta^H)W_{t}N^H_{i,t}+\frac{G_{t}}{N^H}\right]+c_{2}nw^{H}_{i,t}\;\;.
    \end{equation}

    Consumption in \eqref{eq.C} represents the desired spending level for the household. If they are rationed in the goods market, see Section \ref{sec:firms}, then they are left with involuntary saving, which is added to their stock of deposits and CBDC.

    Each household is matched to several banks, see Section \ref{NETTOP}. The amount of net wealth that household $i$ allocates to bank $h$ is given by the fraction $w_i^h$; the weights are constant over time and sum to $1$, $\sum_h w_i^h=1$. 
    A fraction of the deposits in bank $h$ is converted to CBDC according to a rule that depends on the leverage of the bank:	
    \begin{equation}
    \label{EQ_CBDC}
        CBDC^h_{i,t}=\psi(RM_{h,t}) 	w_i^h nw^{H}_{i,t}\;\;.
    \end{equation}
     where $RM_{h,t}$ is a leverage risk measure of the bank $h$ to which the household $i$ is matched as depositor, see Section \ref{sec:banks}:
     \footnote{The leverage ratio is defined as assets divided by equity, in our setting bank's equity is $nw^{B}_{h,t}$, assets are 
     $R_{h,t}+\sum_{j=1}^{N^F}L_{hj,t-1}+\sum_{q=1}^{N^B}I^{l}_{hq,t-1}+B_t^h$. Then  $RM_{h,t}$ is the leverage ratio minus 1.}:

     \begin{equation}
     \label{riskbank}
        RM_{h,t}=\frac{D_{h,t}^B+\sum_{z=1}^{Z}I^b_{zh,t-1}}{nw^{B}_{h,t}}\;\;,
    \end{equation}
    $CBDC^h_{i,t}$ denotes the amount of liquidity withdrawn from bank $h$ by household $i$ and converted into CBDC. The holding of CBDC by household $i$ is 
    $$
    CBDC_{i,t}=\sum_h CBDC^h_{i,t}\;\;.
    $$
    Notice that the allocation between deposits and CBDC concerns the wealth stock and not saving. In Section \ref{sec:results}, we provide a detailed description of the function $\psi(RM_{h,t})$ in (\ref{EQ_CBDC}), which governs the transfer of liquidity to CBDC.
    
    The hypothesis that the substitution of deposits with CBDC is driven by the leverage ratio of the bank is inspired by bank-run models such as \citet{GERT_KYI,veraart2025,feinstein2025}. As discussed in \citet{BIND2}, deposit transfers to safe banks could also play a relevant role; in our model, as in all the literature on bank-run and CBDC, we simplify the analysis assuming that substitution of deposits only involves Central Bank digital money.  
       
    \subsubsection{The labour market}
    \label{sec.labour}
    
    The labour supply is given by the number of households. Each of them provides one unit of labour inelastically, and, therefore, each household corresponds to one worker. As the labour supply is perfectly inelastic, employment is determined by the demand of firms. Workers are homogeneous, they share the same skills and productivity.

    Firms adjust their labour input to their target labour demand defined in (\ref{eq.labour_target}) by hiring or firing workers. 
    Firms and workers are matched at $t=0$, then firms seeking to expand the workforce retain all current employed workers and try to hire additional ones from the pool of unemployed workers. In case of excess aggregate labour demand, available workers are hired proportionally to firms' individual demands. 
    If labour demand falls short of the current number of employed workers, firms fire workers, who become job seekers. Job seekers strive to return to employment by applying for new jobs. 
    
    The matching mechanism regulating the transition from unemployment to employment follows a binomial probability model by which a job seeker can successfully secure a new job within a predetermined number of attempts per unit of time, see Section \ref{par.calibration_labour_mkt}. The same mechanism applies for the matching at $t=0$. The mechanism creates involuntary unemployment as not all job seekers return to employment immediately, preventing the unemployment rate from unrealistically dropping to zero. 
    
     The wage rate $W$ is determined by a representative trade union of all workers. We assume that the wage rate adjusts sluggishly, based on an adaptive mechanism, to prevent the wage from jumping up or down sharply: the wage decreases (increases) when unemployment is above (below) a target unemployment rate $u^{\star}$. The mechanism is a stylized representation of wage dynamics,
    wages move upward when the economy tends to full employment and downward when the unemployment rate is high.
    
    We assume that the wage rate evolves as
 	\begin{equation}\label{eq.Phillips}
	W_t=
	\begin{cases}
	W_{t-1} (1+\gamma^W) & \mbox{if}\ u_t < u^\star \\
	W_{t-1}(1-\gamma^W) & \mbox{if}\ u_t \geq u^\star \\
	\end{cases}\;\;,
	\end{equation}
    where $u_t$ is the unemployment rate and $\gamma^W$ is a random variable uniformly distributed between $0$ and $w_b$. 
			
    \subsection{Government and Central Bank}
    \label{sec.GOV}
    The government collects taxes and CB's profits, issues debt, and distributes lump sum transfers to households.

    Government bonds are acquired by the CB and banks.
    Bonds have a one-period maturity, are issued at par (with unitary face value) and pay an interest rate $r^B$. The government's budget constraint is
	\begin{equation}\label{eq.deltaB}
	\Delta B_{t}= r^{B}B_{t}+G_t -T_{t}-\Pi_{t}^{CB}\;\;,
	\end{equation}
    where $B_t$ is the outstanding stock of government bonds,  $G_t$ denotes public transfers to households, $T_{t}$ denotes tax revenues and $\Pi^{CB}_t$ is the profit of the CB, repatriated to the government.
	
    To concentrate our attention on the effects associated with the introduction of CBDC, we assume a zero balance for the government budget ($\Delta B_t=0$) and, therefore, $G_t$ is such that
    \begin{equation}\label{eq.G}
	G_t=T_{t}+\Pi_{t}^{CB}-r^{B}B_{t}\;\;.        
    \end{equation}

    Note that the transfers to households $G_t$ can be positive or negative depending on whether government interest expenses are below or above revenues. $G_t$ is distributed to all households in the same way ($\frac{G_t}{n^H}$).
    
    The profit of the CB is given by interest payments on bonds (${B}_t^{CB}$) minus remuneration of banks' reserves $R_t$ and of $CBDC_t$:
	\begin{equation}\label{eq.Pi_CB}
	\Pi_{t}^{CB}={B}_{t-1}^{CB}r^{B}-R_{t-1}r^{L}-CBDC_{t-1} r^{CBDC}\;\;.
	\end{equation}
    $CBDC_{t-1}$ denotes the aggregate volume of CBDC at time $t-1$ and $r^{CBDC}$ is its remuneration rate.
    The interest rates on reserves ($r^{L}$) and on CBDC ($r^{CBDC}$) are kept constant over time.

    The CB retains all government bonds except those acquired by commercial banks:
    ${B}_{t}^{CB}=B_t -\sum_h B_t^h$, where $B_t^h$ is the quantity of government bonds owned by bank $h$ which is determined by the rule described in Section \ref{sec:banks}. 
		
    According to the aggregate balance sheet identity for the whole economy, the negative net wealth of the government is balanced by the positive net wealth of the private sector so that aggregate net wealth is zero, see Appendix \ref{sub.agg_bal and trans_mat}  for further details:
		\[\sum_{i=1 }^{N^H}nw^{H}_{i,t}+\sum_{j=1}^{N^F} nw^{F}_{j,t}+\sum_{h=1 }^{N^B}nw^{B}_{h,t}+nw^{G}_t=0\;\;.\]
        
    \subsection{The business sector}\label{sec:firms}
    There are $N^F$ firms, indexed by $j$, producing a homogeneous good using only labour as input.  
    In order to hire workers, firms need to pay the wage bill in advance. This cash-in-advance constraint is binding, so firms can only hire workers up to the available liquidity.
    
    The balance sheet of a firm is made up of bank deposits $D^F$ on the asset side and liabilities consisting of loans $L^F$. The net wealth is provided by:\footnote{ Inventories are perishable as goods are assumed to fully depreciate each period. This assumption rules out business cycles driven by the accumulation of inventory. However, business cycles can arise from variations in business expectations driven by variations in sales. } 
    \begin{equation}\label{eq.nwF}
        nw_{jt}^{F}=D_{jt}^{F}-L_{jt}^{F}\;\;.
    \end{equation}
  
    In each period, firms make their decisions in the following sequence:
    \begin{enumerate}[itemsep=0.05cm]
        \item Set an output target from which they derive the labour target.
        \item Seek financing in order to meet the expected wage bill by borrowing if needed.
        \item Hire workers until the wage bill is met or no further employable workers can be found, then produce.
        \item Set a price for their output and sell it in the market. 
    \end{enumerate}

    We follow \citet[][see Figure 3.1, p.~57]{delli2011macroeconomics}, assuming that firms set target quantities \eqref{eq.Y_adustment} and prices \eqref{eq.mark-up} considering the un-sold production (inventory) and firm specific mark-ups.
    
    Firm $j$'s output target $Y^{target}_{j,t}$ at time $t$ is determined as follows:	
	\begin{equation}\label{eq.Y_adustment}
	Y^{target}_{j,t} =
	\begin{cases}
            Y^s_{j, t-1} (1-\chi^q_j) & \mbox{ if}\ INV_{j,t-1} \geq \gamma_q Y^s_{j,t-1} \mbox{ and}\ P_{j,t-1} \leq \gamma_p \bar{P}_{t-1} \\
            Y^s_{j,t-1}(1+\chi^q_j)  & \mbox{if}\ INV_{j,t-1} < \gamma_q Y^s_{j,t-1} \mbox{ and}\ P_{j,t-1} > \gamma_p \bar{P}_{t-1}
	\end{cases}\;\;,
	\end{equation}
        where $Y^{target}_{j,t}$ is the target output of firm $j$ at time $t$, $Y^s_{j,t-1}$ is the output produced in $t-1$, $Y_{j,t-1}$ is the output sold in $t-1$, and $\chi^q_j \sim U(0,\ q_b)$ is a uniform random variable, $INV_{j,t-1}$ is the inventory in $t-1$, $\gamma_q$ is a calibrated threshold, $P_{j, t-1}$ is the price set by firm $j$, and $\bar{P}_{t-1}$ is the average market price at $t-1$.

    The rule establishes that if a firm succeeds in  selling at least a fraction $1-\gamma_q$ of its output, and the price set is greater than $\gamma_p$ of the market price, then the one-step ahead target is revised to be above the actual output.     
    In case the firm has not sold at least a fraction $1-\gamma_q$ of its output and its price is lower than or equal to $\gamma_p$ of the market price, then the target is revised to be below the actual output. In the remaining two cases, the target is left unchanged.
	
    The firm's labour target directly follows from the output target: 
	\begin{equation}\label{eq.labour_target}
	N^{target}_{j,t} =\frac{1}{\alpha}Y^{target}_{j,t} \;\;,
	\end{equation}
    where $\alpha$ is labour productivity, each worker produces $\alpha$ units of goods.

    The main issue for the firm is whether and how this labour demand is financed under the cash-in-advance constraint. 
    On the basis of the target output and employment, each firm computes a target wage bill and tries to ensure the liquidity to finance it. To this end, the company uses first its own available resources and then goes to the credit market to borrow any additional need. 
    Therefore, the loan target is
    \begin{equation}\label{eq.loantarget}
	L^{target}_{j,t} = \max(0,\; W_{t} N^{target}_{j,t}- \zeta nw^F_{j,t})\;\;,
    \end{equation}
    where $nw^F$ is the net wealth of the firm. $\zeta \in [0,1]$ is a parameter that weighs the relative priority given to internal finance ($\zeta=1$) over borrowing ($\zeta = 0$) to meet operational needs. 
	
    As a firm might be rationed in the credit market, its actual loan $L_{j,t}$ might be smaller than its loan target
	\[
	L_{j,t} \leq L^{target}_{j,t} \;\;.
	\]
    In Section \ref{sec:banks}, we detail how the credit market works and, therefore, why not all the demand for credit may be fulfilled.	Once the loan has been obtained, it is added to the firm's deposit account and the funds are immediately available. The loans last only one period. 

    Once the firm has secured a loan and updated its liquidity, it determines the expected wage bill $\Omega$  by balancing its target output against the available funds:
    \begin{equation}\label{wagebill} 
	\Omega_{j,t}  = \mbox{min}\left[D^F_{j,t}, W_t N^{target}_{j,t}\right]\;\;.
	\end{equation}
	As described in Section \ref{sec.labour}, in case the labour  supply is lower than the  demand, the  firm can hire workers proportionally to its demand compared to the total demand. 
    As a consequence, the number of employed people $N_{j,t}$ of firm $j$ satisfies the condition
	\[
	N_{j,t} \leq  \frac{\Omega_{j,t}}{W_t}.  
	\] 
  and the actual supply of the firm is 
	
    \begin{equation}\label{eq:prod_funct}
    Y_{j,t}^{s}=\alpha N_{j,t}.
    \end{equation}
	
    The price of the goods produced by firm $j$ is determined as a mark-up $\mu_{j,t}$ on the unit cost, due to its monopolist power:
    
        \begin{equation}\label{eq.price}
    	P_{j,t}=(1+\mu_{j,t})uc_{j,t}\;\;.
	\end{equation}

    The cost of producing one unit of good is $uc_{j,t}$  and is defined as the ratio of the wage bill plus the cost of borrowing to the actual output of the company: 
	\begin{equation}\label{eq.uc}
    	uc_{j,t}=\frac{(W_t N_{j,t} + r^f_{j,t}L_{j,t})}{Y^s_{j,t}}\;\;,
	\end{equation} 
    where $L$ represents the total amount of bank borrowing and $r^f$ is the associated interest rate.

    Similarly to quantity adjustment, firms revise their mark-ups after observing the inventory-to-production and price-to-market price ratio.
    The mark-up of a firm is bounded from above and below and it goes up (or down) depending on the inventory being below (above) a certain threshold $\gamma_q$ of production and past price being below (above) a threshold $\gamma_p$ of the market price. In the remaining two cases, the mark-up doesn't change. Specifically, the mark-up charged by firm $j$ at time $t$ follows the rule
	\begin{equation}\label{eq.mark-up}
	\mu_{j,t} =
	\begin{cases}
    	\min[\mu_{max},\mu_{j,t-1}(1+\chi^{\mu}_j)] & \mbox{ if}\ INV_{j,t-1} \leq \gamma_q Y^s_{j,t-1} \mbox{ and}\ P_{j,t-1} \leq \gamma_p \bar{P}_{t-1}\\
	    \max[\mu_{min},\mu_{j,t-1}(1-\chi^{\mu}_j)]  & \mbox{ if}\ INV_{j,t-1} > \gamma_q Y^s_{j,t-1} \mbox{ and}\ P_{j,t-1} > \gamma_p \bar{P}_{t-1}\
	\end{cases}\;\;,
	\end{equation}
    where $\chi^{\mu}_j \sim U(0,\ \mu_b)$ is a uniform random variable.
    
    After production and pricing have been determined, the goods market opens and consumers spend their consumption budget $C^d_{i,t}$ in (\ref{eq.C}) following the matching mechanism described in \citet{gurgone2018effects}.

    Rationing can occur in the goods market and, therefore, actual sales can be smaller than the actual output ($Y_{j,t} \le Y^s_{j,t}$).

    Given the output $Y_{j,t}$ sold by firm $j$, the firm's gross profits $\Pi^F_{j,t}$ are
    \begin{equation}\label{eq.profits_firms}
        \Pi_{j,t}^F=P_{j,t}{Y_{j,t}}-W_tN_{j,t}+D_{j,t-1}^{F}r^{D} - \sum_{h=1}^{N^b}r^f_{jh,t-1}L_{jh,t-1}\;\;.
    \end{equation}
    Gross profits are given by sales revenues minus wage costs (associated with the actual output) and interest charges (deposits and loans). 
 
     If $\Pi^{F}_{j,t}>0$, then the firm pays taxes and dividends, otherwise it absorbs the losses through its bank deposits. 
    If the gross profits are positive, then net profits are given by gross profits minus taxes imposed at the rate $\theta^F$. 
 
    A share $\delta^f_t$ of net profits is distributed as dividends. The share is made up of two parts: a fixed component $\delta^F$ and a component that depends on the net wealth of the firm relative to its after-tax profits that depends on an additional parameter $d^F$: 
	\[
	\delta^f_t = \delta^F + d^F \frac{nw^F_{j,t}}{(1-\theta^F)\Pi^F_{j,t}} \;\;.
	\] 
    This dividend policy prevents firms becoming too large on the basis of retained earnings. As a matter of fact, dividends increase (decrease) when net wealth goes up (falls). 
    The firm's net wealth evolves as follows
    \begin{equation}
    nw^{F}_{j,t} =
    \begin{cases}
	(1 - d^F) nw_{j,t-1}^{F}+(1-\theta^F)(1-\delta^F)\Pi_{j,t-1}^{F} \quad \quad &\text{if} \quad \quad \Pi_{j,t-1}^{F}>0\\
    nw_{j,t-1}^{F}+\Pi_{j,t-1}^{F} \quad \quad &\text{if} \quad \quad \Pi_{j,t-1}^{F}\leq 0
    \end{cases}\;\;.
    \label{eq.nw_F_dynamics}
    \end{equation}
    If $nw_{j,t}^{F}\geq 0$, then the firm's debt is serviced, otherwise the firm becomes insolvent and bankruptcy occurs.

    From $t$ to $t+1$ the outgoings of the company consist of wage payments, taxes, dividends, and interest payments on the loan. These payments are settled at the end of period $t$. 
		
    \subsection{The banking sector} 
    \label{sec:banks}

    There are $N^B$ banks, indexed by $h$. They fund themselves through short-term unsecured liabilities and extend loans to firms. In the event of excess liquidity or shortages, they either exchange liquidity in the interbank market or, if they are rationed in that market, sell assets to adjust their liquidity position. 

    \subsubsection{Balance sheet}\label{sec.banks_balance_sheet}
    The asset side of the balance sheet of banks includes outstanding loans to firms, indexed by $j$ and to banks, indexed by $q$, denoted respectively by $L$ and $I^l$, plus reserves $R$ detained at the CB  and government bonds $B$\footnote{$R_h$ is the total reserves held at the CB. It includes the required reserves computed using a constant regulatory reserve ratio $rr$ applied to total deposits of banks, $rr D^B_h$, and any excess reserves, $R_h - rr D^B_h$.}. Liabilities include interbank borrowing $I^b$ from other banks, indexed by $z$, and deposits $D^B$. Bank $h$'s net wealth is given by:

    \begin{equation}\label{eq.nwB}
	nw_{h,t}^{B}=R_{h,t}+\sum_{j=1}^{N^F}L_{hj,t-1}+\sum_{q=1}^{N^B}I^{l}_{hq,t-1}+B_{h,t}-D_{h,t}^B-\sum_{z=1}^{N^B}I^b_{zh,t-1}\;\;.
    \end{equation}
    Banks buy government bonds proportionally to their deposits: $B_{h,t}=\phi_{ext}D_{h,t}$, where we set $\phi_{ext}=10\%$. Without prejudice to the main results, we assume that the bank cannot get liquidity from the CB.

    \subsubsection{Credit supply}\label{sec.credit_supply}
    At the beginning of each period, banks face credit requests from firms and try to serve them in full, while respecting regulatory constraints and internal risk management standards.
    Prudential regulation imposes minimum capital requirements, by which the net wealth of a bank should be greater or equal than a fraction $1/\lambda$ of the risk-weighted assets (RWA) computed according to the Standard approach in the spirit of Basel II and III regulation:	
    $$
    nw^B_{h,t} \geq \frac{1}{\lambda}\text{RWA}_{h,t}\;\;,
    $$
    where $RWA_{h} = \omega_1 \sum L_h + \omega_2 \sum I^l_h$, while the weight for cash and bonds is set at $0$.

    Moreover, we assume that banks hedge against risk in their exposures by keeping a level of net wealth that is able to absorb potential losses under a worst-case scenario. Losses in the worst-case scenario are assessed as a time-varying fraction of total exposures
    $$ nw^B_{h,t} \geq VaR^{tail}_{h,t} (L_{h,t} + I^l_{h,t})\;\;,$$
    where $VaR^{tail}_{h,t}$ is computed as a parametric Gaussian Value at Risk (VaR) at tail probability $tail = 0.99$ where the mean and variance parameters are estimated from the historical losses of the loan portfolio. 

    These two constraints jointly determine the total bank credit supply in the model. Accordingly, bank $h$ supplies credit up to the lowest amount allowed by prudential regulation and portfolio-risk management:\footnote{The expression for the credit supply in \eqref{eq:loan_supply} is obtained by solving $nw^B \geq \frac{1}{\lambda}(\omega_1 L + \omega_2 I^l)$ and $nw^B \geq VaR^{tail}(L + I^l)$ for $L$, where the loan maturity is one time unit.}

    \begin{equation}
    \label{eq:loan_supply}
        L^s_{h,t} = \min \left(
        \lambda \frac{nw^B_{h,t}}{\omega_1} - \omega_2 \frac{I^l_{h,t}}{\omega_1}, 
        \frac{nw^B_{h,t}}{VaR^{tail}_{h,t}} - I^l_{h,t}
        \right)\;\;.
    \end{equation}

    Besides complying with (\ref{eq:loan_supply}), banks manage firm-specific risk so that the exposure to a single firm $j$ is capped at $\varsigma=0.15$ of the bank net wealth, which determines the maximum of equity loss the bank is willing to bear per loan.
    Therefore, the credit supplied to firm $j$ by bank $h$ is up to
    \begin{equation}
    \label{eq:loan_exposure}
        L^s_{hj,t} \leq \frac{\varsigma nw^B_{h,t}}{\rho^f_{j,t}}\;\;,
    \end{equation}
    where $\rho^f$ is the probability of default of firm $j$, that we will define in (\ref{eq.rho_f}). 

    \subsubsection{Banks-firms matching} 
    Firms are matched to banks in the credit market via a preferential attachment mechanism with probabilistic switching, by which firms can switch between lenders with a predetermined probability, see also Section \ref{sec.matching}.
    Each bank charges an interest rate, taking into account its counterparty credit risk and its own cost of funding, leading to heterogeneous interest rates. 
    Banks rank firms in ascending order based on their credit merit.
    They then begin by fully satisfying the loan requests of the least risky firms up to the constraint (\ref{eq:loan_exposure}), continuing in this manner until their total credit supply in \eqref{eq:loan_supply} or demand is exhausted. In this way, risky firms are more likely to be rationed. If a firm is rationed by its preferred bank, then it can seek credit from some other banks, repeating this process until its demand is fully met or all connected banks deny the loan.

    \subsubsection{Interest rate}	
    The default risk $\rho^f_{t,hj}$ perceived by bank $h$ concerning firm $j$ is inspired by \citet{delli2011macroeconomics} and is given by: 
	\begin{equation}\label{eq.rho_f}
	\rho^f_{t,hj} = v_0e^{v_1\left(\frac{l_j}{l\star}-1\right)}\;\;,
	\end{equation}
    where $l_j$ is the loan demand to net wealth ratio of $j$, $v_0$, and $v_1$, $l^\star$ are parameters to be calibrated. 

    The interest rate at which bank $h$ offers a loan to firm $j$ is denoted by $r_{t,hj}^{f}$. The rate is a function of its cost of funding and $j$'s specific risk of default:
    \begin{equation}\label{eq.rflow}
	r^{f}_{hj,t}=\frac{1+cf_{h,t}}{1-\rho^f_{j,t}}-1\;\;,
    \end{equation}
    where $cf_{h,t}$ is the bank $h's$ cost of funds, and is defined as:
    \begin{equation}\label{eq.cf}
	cf_{h,t}=\omega^{D}_{h,t}r^{D}+\omega^{I}_{h,t}{r}_{t-1,h}^{b}\;\;,
    \end{equation}
    where $\omega_{h,t}^i$ represents the share of each source of liquidity of the bank ($i$ represents deposits and interbank borrowing) over liabilities.
    
    By design, the lending rate in \eqref{eq.rflow} is set to be greater than or equal to the bank's funding cost and increases with the probability of default of the firm $j$.

    \subsubsection{Interbank market}\label{sec.ib_market}
    Banks mitigate the risk of illiquidity by trading on the interbank market.  
    We slightly depart from the original framework in \citet{gurgone2018effects,gurgone2022macroprudential}  by limiting the maximum amount that a bank can borrow to the risk-weighted value of its assets, which provides a collateralization of interbank borrowing. 

    Banks aim to set aside a sufficient buffer of liquidity to hedge against changes in the balance sheets of other agents, e.g. withdrawals and defaults.
    This approach results in banks forming an internal liquidity coverage ratio which determines demand and supply of interbank funds.\footnote{Banks may alternatively participate in the interbank market to meet the prudential liquidity coverage ratio mandated by Basel III as in \citet{lilit2016taming}.}

    To enhance realism in terms of interlocked balance sheets, we assume that there are three sessions of the interbank market for each time iteration: i) after lending to firms, ii) after firms sell their production on the goods market, iii) after factoring defaults and losses of firms and other banks. At the end of each session, banks settle their positions. If a bank does not satisfy its liquidity coverage ratio, it liquidates part of its assets through the mechanism in \eqref{eq:ass.price}.

    In agreement with the regulation, bank $h$ computes a liquidity coverage ratio given by
    \begin{equation}\label{eq.LCR}
        LR_{h,t} = \frac{R_{h,t} - rrD_{h,t}^B}{out^E_{h,t} - in^E_{h,t}} \geq 1\;\;,
        \end{equation}
        where the numerator captures the bank’s reserves detained at the CB, adjusted for the required reserve ratio on deposits.  The denominator represents the expected gap between expected cash outflows $out^E$ and inflows $in^E$ over a single period of time.\footnote{As banks anticipate their liquidity requirements based on economic conditions, shown by borrowers' default risks, they require a greater cash level during periods of substantial losses compared to stable times.}
        For the bank to be considered sufficiently liquid, the liquidity ratio must be greater than one. 
        The expected cash outflows consist of interest payments on deposits and interbank borrowing $cib^E$, and the expected amount of lending to firms $L^E$:
        \[ 
        out^E_{h,t}=r^D D_{h,t}^B + cib^E_{h,t}  + L^E_{h,t}\;\;, 
        \] 
        where $L^E_{h,t} = a^L L_{h,t} + (1-a^L) L^E_{h,t-1}$ and $cib^E$ is constructed as the weighted average of the interest rate on interbank borrowing for the borrowed amount during the last $\tau$ periods in the bank's memory. 

    The expected cash inflows $in^E$ include the total amount of interest payments on loans to firms from the subset of borrowers $J$, along with the principal amount to be repaid at the end of period $t$, adjusted for the probability of default of borrowers, together with the interest paid by the CB on reserves and bonds:
    \[
    in^E_{h,t}=\sum_{j\in J}L_{hj,t} (r^f_{hj,t-1} + 1-\rho^f_{hj,t}) +r^L R_{h,t} + r^B B_{h,t} \;.
    \]
    If the liquidity coverage ratio is lower than one, meaning that the expected cash outflows net of inflows are greater than bank reserves net of required reserves, then bank $h$ demands interbank liquidity ($I^d$) to close the gap.

    \begin{equation}
        I^d_{h,t} = out^E_{t,h} - in^E_{t,h}- (R_{h,t} - rrD_{h,t}^B)\;\;.
    \end{equation}\label{eq.Id}
    Otherwise, the bank supplies its excess liquidity ($I^s$) on the interbank market, subject to its total loan supply in \eqref{eq:loan_supply} net of outstanding loans:

    \begin{equation}
        I^s_{h,t}=\min\left[\vphantom{\sum_{a}^{b}} R_{h,t}-rrD_{h,t}^B-(out^E_{t,h} - in^E_{t,h}),\ L^s_{h,t} - \sum_{j\in J}L^F_{hj,t-k}  \right]\;\;.
    \end{equation}\label{eq.Is}

    Furthermore, we assume that the liquidity supplied by a bank $h$ to any bank $z$ cannot  exceed the value of the illiquid assets of $z$: 
    \begin{equation}
        I^s_{hz,t} \leq \sum_{j\in J}L_{zj,t} (1-\rho^f_{zj,t}) + B_{z,t}\;\;,
        \end{equation}
    where $J$ is the subset of firms that borrow from the bank $z$ and $\rho^f$ is the default probability of these firms.

    Banks trade in a decentralized interbank market. 
    Banks in demand of liquidity enter one-by-one in a random order and are matched to a randomly selected bank offering a positive supply of interbank funds. This process is repeated $n^{ibtent}$ times for all potential borrowers.
    Borrowers place a bid $r^{bid}$ reflecting how much they are willing to pay on borrowed funds:
    \begin{equation}
        r_{z,t}^{\text{bid}} = \frac{r^H + r^L}{2} (1 + \varepsilon_{z,t}), \quad r_{z,t}^{\text{bid}} \in [r^L, r^H]\;\;.
    \label{eq:bid_rate}
    \end{equation}
    Since they do not know lenders' reservation rates, borrowers initially bid in the middle of the corridor determined by lower and upper-bounds, which are set by the CB and are respectively the rate paid on excess funds $r^L$ (lower-bound) and the rate of a fictitious marginal lending facility $r^H$ (higher bound).\footnote{Although the interest rate on interbank loans is set around the mid-corridor, this model assumes that banks cannot access the marginal lending facility. As a result, unmet funding needs are covered by the liquidation of assets.}

    If demand is not entirely satisfied, the borrowers increase the bid by a mark-up $\varepsilon$, whereas they decrease it by the same amount if a lender accepts their bid:
    \begin{equation}
        \varepsilon_{z, o+1} = 
    \begin{cases} 
    \varepsilon_{z, o} + \gamma^{ib} & \text{if } I^d_{z, o} > I^b_{z, o} \quad \text{and} \quad r^{bid}_{z, o} \leq r^H \\ 
    \varepsilon_{z, o} - \gamma^{ib} & \text{if } I^d_{z, o} = I^b_{z, o} \quad \text{and} \quad r^{bid}_{z, o} \geq r^L 
    \end{cases}\;\;,
    \label{eq:ib_mkup}
    \end{equation}
    where $o \in[1,n^{ibtent}]$ denotes the sequence of borrowing attempts and $\gamma^{ib}\sim U(0,\ \text{bid}_b)$ is a uniform random variable.
    It should be noted that the last value of $\varepsilon$ is retained in the bank's memory at the beginning of a new interbank session.

    The reservation rate of lenders $r^{res}$ is determined by a risk premium for the probability of default of borrowers relative to the rate on excess funds $r^L$. For a lender $h$ and a borrower $z$, it is defined as
    \begin{equation}\label{eq.rb}
        r_{hz,t}^{res}=\frac{1+r^{L}}{1-\rho^b_{hz,t}}-1\;\;,
   \end{equation}
   where the default probability
   \begin{equation}
        \rho_{z,t}^{b} = v_0 \exp \left[ v_1 \left( \frac{l^b_{z}}{l^{b\star}} - 1 \right) \right]
        \label{eq:def_prob_ib}
    \end{equation}
    grows with $l^b_z$, the total exposures to equity ratio of bank $z$. $v_0$, $v_1$, and $l^{b\star}$ are parameters to be calibrated. 

    The amount borrowed (lent) at time $t$ by $z$ ($h$) is $I_{hz,t} = \min(I^d_{z,t},I^s_{hz,t})$ and  takes place at an interest rate where the borrower's bid rate is greater or equal than the lender's reservation rate:
    \begin{equation}
        r^b_{hz,t} = r^{bid}_{z,t} \quad\mbox{if }\;\; r^{bid}_{z,t} \geq r^{res}_{hz,t}.
    \end{equation}

    \subsubsection{Liquidation of assets}\label{sec.liquidation}
    When banks run out of liquidity or need to settle creditors' claims following bankruptcy, they sell off assets (government bonds and loans to firms). The liquidation process is handled by a special agency. The index $o = 1, \dots, N^B$ tracks the order of sellers within each time unit, as more than one bank can sell assets at time $t$. Sellers enter the market in random order: the first one sells at the most favourable price $p_1 > p_2$ per unit of bond or loan, the second at $p_2 > p_3$ and so on. At the end of period $t$, the asset price is reset to its initial value $p_0 = 1$. As in \citet{SHIN,veraart2025,feinstein2025}, we assume that banks begin by selling the most liquid assets, i.e., government bonds, that are more liquid than loans ($\epsilon^{bonds} >  \epsilon^{loans}$), and stop when their liquidity needs are met. The special agency  purchases assets at the price $p_{o}$:

    \begin{equation}
    \label{eq:ass.price}
        p^i_{o}=\max\left[0.5,\; p^i_{o-1}\left( 1-\frac{ q^i_{o}}{Q^i} \frac{1}{\epsilon^i} \right)\right], \quad\quad i = \{bonds, \ loans\}\;\;,
    \end{equation}
    where $q^i_{o}$ is the amount that a bank in rank $o$ needs to liquidate, $Q^i$ is the total amount of asset $i$ in the economy, and $\epsilon^i$ is the price elasticity of asset $i$.
    The lower-bound on $p$ is set to $0.5$ to reflect the upper-bound on the loss given default of assets and the role of arbitrageurs, who would purchase underpriced assets at a discount to make profits, thus preventing prices from dropping further.
    The assets acquired by the agency are held to maturity and any profit or loss is transferred to the government. Liquidation has a price impact only on the balance sheet of the seller. 
    The lower-bound can also be interpreted as a bail-out of banks in very difficult conditions.

    \subsubsection{Profits and losses}\label{par.P&L}
    Gross profits for bank $h$ are 
    \begin{align}\label{eq.PiB}
	\Pi_{h,t}^{B}=&R_{h,t-1}r^{L}+B_{h,t-1}r^B+\sum_{j=1}^{N^F}{L}_{hj,t-1}r_{hj,t-1}^{f} + \nonumber \\
	&+\sum_{q=1}^{N^B}I^l_{hq,t-1}r^b_{hq,t-1}-\sum_{z=1}^{N^B}I^b_{zh,t-1}r^b_{zh,t-1}-D_{h,t-1}r^{D} - losses_{h,t}\;\;.
    \end{align}
    
    If positive, profits are taxed at rate $\theta^B$ and a share $\delta^B$ is distributed to shareholders, what is left is retained by the bank.

    Losses arise from three different sources: defaults on loans to firms, defaults on interbank loans, and liquidation of assets:
    \begin{itemize}
    \item Losses from defaults on loans to firms are given by the stock of loans outstanding to insolvent firms, minus their deposits, which are seized in case of default.  In other words, losses are provided by the negative net wealth of firms defaulting on loans. If an insolvent firm borrowed from more than one bank, then the loss borne by creditors is distributed proportionally to the amount lent by each bank.
    \item Banks defaulting on interbank loans are another potential source of losses. If a bank defaults, then creditors recover their share of residual assets in proportion to their claims on the defaulter's liabilities.
    As claimants include households, firms, and other banks, we assume that under bankruptcy law, depositors (households and firms) are the most guaranteed type of creditors. Their claims are therefore prioritized over interbank claims, which are settled on any residual asset after depositors have been repaid. 
    Notice that contagion can arise. If a borrower defaults, then the creditor bank can become insolvent and go into bankruptcy as well, triggering a cascade of bankruptcies or losses on the interbank and credit market.
    \item Losses from asset liquidation are provided by the difference between net wealth before and after liquidation.
    \end{itemize}

    At the end of each period, the net wealth of bank $h$ is updated by net profits: the bank retains profits after taxes, distributes dividends if gross profits are positive and absorbs losses otherwise:

    \begin{align}\label{eq.deltanwB}
        nw_{h,t}^{B} =
        \begin{cases}
            nw_{h,t-1}^{B} + (1-\theta^B)(1 - \delta^B)\Pi_{h,t}^{B}, & \text{if } \Pi_{h,t}^{B}>0 \\ 
            nw_{h,t-1}^{B} + \Pi_{h,t}^{B} & \text{if } \Pi_{h,t}^{B}\leq 0 \\ 
        \end{cases}\;\;.
    \end{align}

    The end-of-period change in reserves held by bank $h$  with the CB is
    \begin{equation}\label{eq.deltaR}
	\Delta R_{h,t} = \Delta D^B_{h,t} - \Delta L_{h,t}+ \Delta nw_{h,t}^{B}
    \end{equation}
    where $\Delta$ is the first difference operator.

    \subsection{Bankruptcy and new entrants}\label{sec.defaults}
	
    If a firm or bank’s net wealth becomes negative, it goes bankrupt. The resulting losses are absorbed by its creditors' balance sheets, potentially triggering further defaults. 

    Households act as shareholders of both firms and banks, receiving dividends as part of their participation in profits. For simplicity, we assume that each firm or bank is equally owned by a fixed number of households, who are equivalent to depositors in the case of banks.

    \subsubsection{Firms}
    Banks do not lose the entire loan amount when a firm defaults on its loans. Instead, the firm's remaining assets (which are equivalent to deposits) are distributed among creditors. The actual loss corresponds to the firm's negative net wealth which is shared proportionally among creditors.
    
    Consequently, if a defaulting firm has more than one creditor, the actual loss is distributed across all creditors, with each creditor bearing a loss proportional to the size of its loan relative to the borrower's total debt. After default, firms exit the market and are replaced after  $recap^F=2$ periods by new firms. These entrants begin with no liabilities and positive deposits, funded by a randomly determined share of their shareholders' wealth. 

    \subsubsection{Banks}
    A bank in default typically has multiple creditors, as its liabilities include deposits from firms and households, as well as interbank loans. In the event of default, the bank’s creditors absorb its negative net wealth until it is fully depleted. All creditors incur losses; however, depositors — especially households — are the most protected. They only bear the portion of losses not already absorbed by other creditors, as they rank last in the loss hierarchy. After default, the only remaining items on the bank's balance sheet are deposits and a corresponding amount of reserves. Households and firms retain access to their deposits even if the bank is no longer active and they can still use them for consumption and to pay wages. The bank is not replaced by another institution but is instead recapitalized with fresh capital from its shareholders, either after a minimum period $recap^B=4$ out of operation or once it becomes viable for recapitalization.

     The recapitalization is deemed successful only if shareholders possess enough capital to satisfy the required asset-to-liability ratio; otherwise, the bank remains inactive until its shareholders can finance the operations.

    It is important to note that a bank's default may trigger the default of its creditors, which include firms and other banks. Households are exposed only through their deposits and, in the worst-case scenario, may lose their entire net wealth. If a firm loses part of its deposits, it may become unable to repay its loans and consequently goes bankrupt. A similar mechanism applies to banks, whose balance sheets contain interbank loans that can transmit financial distress throughout the system.

    \section{CBDC adoption}
    \label{ADOP}

    We set different rules for the portion of liquidity held by household $i$ in bank $h$ to be converted in CBDC. We assume that  
    $$CBDC_{ih,t}=\psi(RM_{h,t}) w_{ih} nw^{H}_{i,t}\;\;,$$
    where $RM_h$, as defined in (\ref{riskbank}), is the riskiness of bank $h$ (bank liabilities/net wealth ratio) to which the household is matched. 
    Households are matched to more than one bank and therefore the total amount of CBDC of household $i$ is 
    $$CBDC_{i,t}=\sum_h CBDC_{ih,t}.$$
    Our hypothesis is that banks' balance sheets are observable and households base their conversion decision on their riskiness which is proxied by the leverage ratio. 
    \begin{figure}
        \centering
        \includegraphics[width=0.95\linewidth]{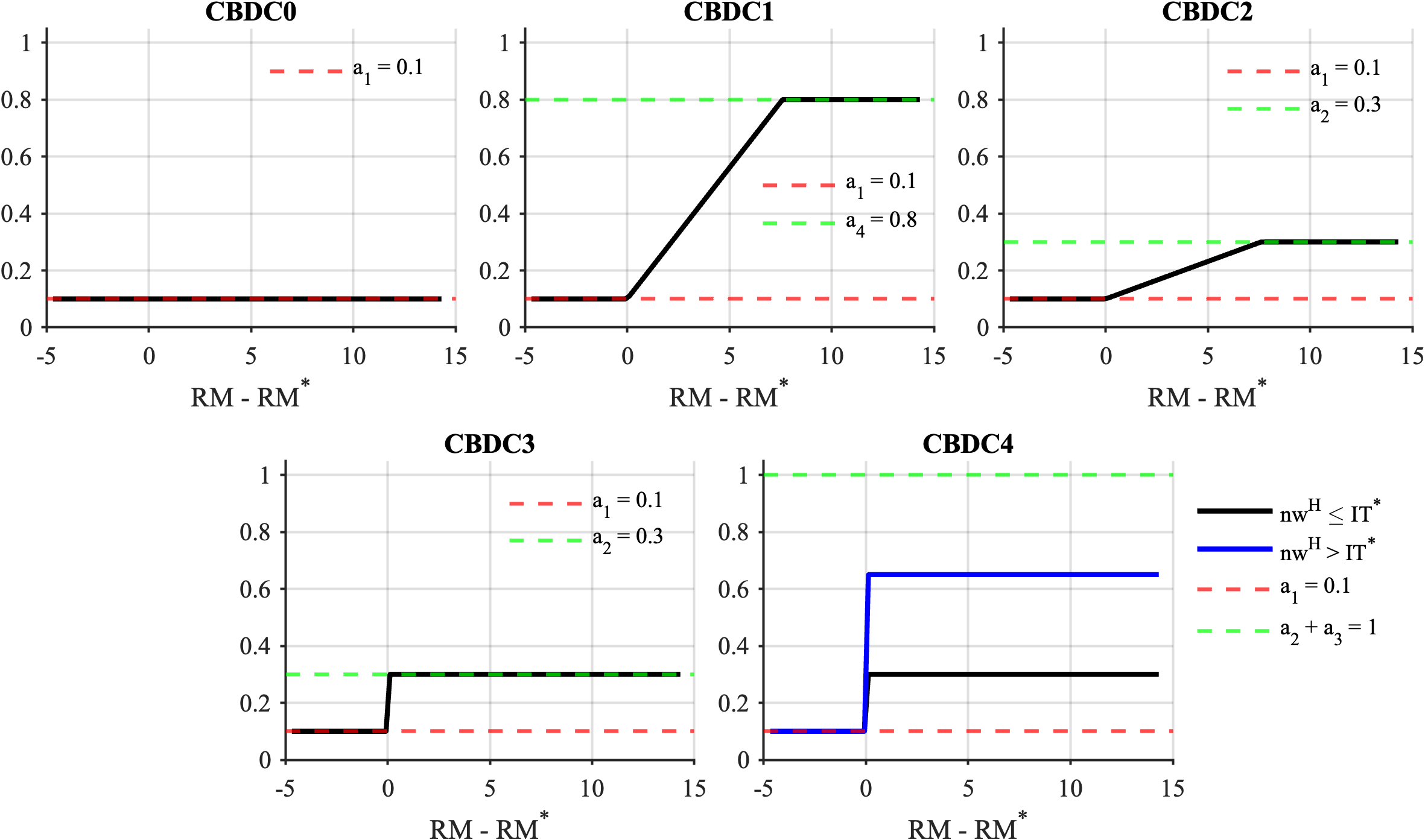}
        \caption{Share of households' liquidity converted into CBDC by adoption rules.}
        \label{fig.cbdc_rules}
    \end{figure}

    We consider five different rules determining the fraction of liquidity to be converted in CBDC: flat fraction (CBDC0),  loose bound on deposit withdrawal dependent on bank's riskiness (CBDC1), fraction dependent on bank's riskiness with a tight bound (CBDC2 and 3), fraction dependent on bank's riskiness and deposit insurance scheme (CBDC4).
    
    The rules are as follows, see Figure \ref{fig.cbdc_rules} for a graphical representation:
    \begin{itemize}
        \item \textbf{CBDC0}:  Households convert $a_1$ of their liquidity into CBDC independently of the riskiness of the bank:
      \begin{equation}\label{eq:cbdc0}
         \psi(RM_{h,t})= a_1.
     \end{equation}
     In the simulations we set $a_1=0.1$.
     \item 
     \textbf{CBDC1}: 
     Households convert a fixed quota of their liquidity into CBDC if the riskiness of the bank is below $RM^*$ and an increasing linear function of the riskiness if it is above $RM^*$. 
     The bound on conversion is loose, the upper-bound is set at $a_4=80\%$  of deposits if bank's  riskiness is above $RM^*+RM^{lim}$ (almost unconstrained substitution):

    \begin{equation}\label{eq:cbdc1}
    \psi(RM_{h,t}) =
        \begin{cases}
        a_1 & RM_{h,t} \le RM^* \\
        a_1 + (a_4-a_1)\dfrac{RM_{h,t}-RM^*}{RM^{lim}} & RM^* < RM_{h,t} < RM^* + RM^{lim} \\
        a_4 & RM_{h,t} \ge RM^* + RM^{lim}.
        \end{cases}
    \end{equation}
    Coherently with the regulation, the risk measure threshold is $RM^*=6$ and $RM^{lim}=7.6$.

    \item 
    \textbf{CBDC2}: Households convert a fraction of liquidity into CBDC depending on the riskiness of the bank.  
    The quota converted into CBDC is up to a cap $a_2 < a_4$: 

         \begin{equation}\label{eq:cbdc2}
     \psi(RM_{h,t}) =
        \begin{cases}
        a_1 & RM_{h,t} \leq RM^* \\
        a_1 + (a_2-a_1)\dfrac{RM_{h,t}-RM^*}{RM^{lim}} & RM^* < RM_{h,t} < RM^* + RM^{lim} \\
        a_2 & RM_{h,t} \geq RM^* + RM^{lim}.
        \end{cases}
    \end{equation} 
    In what follows, we set $a_2=0.3$, that is, depositors convert at most $30\%$ of their deposits into CBDC. 
         
    \item 
    \textbf{CBDC3}: Households convert a fixed quota of their liquidity into CBDC, depending on bank's riskiness. If the risk threshold $RM^*$ is exceeded, households convert the fraction $a_2$ of wealth, otherwise they convert the fraction $a_1$, with $a_1 < a_2 <a_4$: 

    \begin{equation}\label{eq:cbdc3}
    \psi(RM_{h,t}) =
        \begin{cases}
        a_1, & RM_{h,t} \leq RM^* \\[6pt]
        a_2, & RM_{h,t} > RM^*.
        \end{cases}
    \end{equation}
            
    \item 
    \textbf{CBDC4}: Households convert a fixed quota of liquidity into CBDC, the quota depends on the riskiness of the bank. If riskiness is below $RM^*$, then the quota is $a_1$; if it is above $RM^*$ and deposits are below the insurance deposit threshold ($IT^*$), then the quota is $a_2 \ge a_1$; if it is above $RM^*$ and deposits are larger than $IT^*$, then the quota is 
    $a_2$ plus a linear term that depends upon $\frac{w_{ih}nw^H_i - IT^*}{w_{ih}nw^H_i}$:  
 
    \begin{align}\label{eq:cbdc4}
        \psi(RM_{h,t}) =
        \begin{cases}
            a_1 & \text{if } RM_{h,t} \leq RM^* \\ 
            a_2, & \text{if } RM_{h,t} > RM^*\, \mbox{ and } \,w_{ih}nw^H_i \leq IT^* \\ 
            a_2 + a_3 \frac{w_{ih}nw^H_i - IT^*}{w_{ih}nw^H_i}, & \text{if } RM_{h,t} >RM^*\, \mbox{ and } \,w_{ih}nw^H_i > IT^*,
        \end{cases}
    \end{align}
    where $a_2+a_3=1$. Insurance is provided by the government and is funded by public funds. 
    Conditional on the default of bank $h$, the government compensates depositor $i$ according to the following scheme:
    $$
    compensation_{i|default_h} = \min(w_{ih}nw^H_i,\; IT^*) \cdot  \left[1 - \psi(RM_{h,t}) \right] \cdot (1 - \xi^{rec}_{i,h})\;\;,
    $$ 
    where $\xi^{rec}_{i,h}$ is the recovery rate associated to the default of bank $h$.\footnote{The recovery rate follows from the bankruptcy law, by which depositors are the most guaranteed type of creditors, see Section \ref{par.P&L} for further details.} 
      \end{itemize}

    CBDC0 and CBDC1 rules describe the two central scenarios of our analysis.
    In the first, households hold a fixed amount of their liquidity  as CBDC (10\%), the amount is calibrated to be similar to the amount of cash. In the second scenario, we consider the extreme case in which a massive conversion of deposits into CBDC can occur depending on bank's riskiness. This scenario allows for a bank-run to be ignited by the conversion of deposits into CBDC when default of the bank is feared. 

    CBDC2 and CBDC3 rules allow us to evaluate the effects of tight bounds on the amount of CBDC with households switching from $10$ to $30\%$ in case the bank becomes risky (its leverage triggers a certain threshold); CBDC2 rule allows for a smooth conversion, CBDC3 rule considers an abrupt conversion. 
   The CBDC4 rule models deposit insurance: households substitute deposits with CBDC in a limited way if the bank is risky and deposits are below a certain threshold and substitution goes up if deposits are above the deposit insurance threshold ($100,000$ euro in the Euro area).

    \section{Simulation results}
    \label{sec:results}

    The model is calibrated as discussed in Appendix \ref{sec.calibration} while the Bayesian estimation procedure is reported in Appendix \ref{sec.estimation}. In short, the calibration relies on macroeconomic and financial data from the European Union (EU) and the Euro area. To ensure computational tractability, the number of agents is scaled down from real-world statistics, resulting in 500 firms, 2,500 households, and 10 banks preserving ratios observed empirically. 
    The initial values of net wealth are derived from the deposit-to-GDP ratio reported by the European Central Bank. The nominal wage serves as a numéraire to scale monetary quantities, and the unemployment rate is calibrated to the long-term Euro area average. Transition in the labour market is modeled using a binomial distribution which is matched to observed EU labour market flows. 
    The Bayesian estimation targets the first two moments (mean and standard deviation) of key economic time series (change in consumer prices, credit to GDP, unemployment rate, and CET1 ratio capital of banks) to align simulated to real data. Parameters are calibrated to closely match the mean of macroeconomic variables minimizing the variability. Table \ref{tab.estimation_targets} reports the time series and target moments for the 2000-2019 period (except CET1 ratio for 2015-2019).

    Once the model is calibrated, we analyze the impact of introducing a CBDC through simulations. Each simulation runs for 1,000 time steps, but only the last 500 iterations are considered to avoid the influence of initial conditions. To ensure comparability, the network structure and the random seed are held constant across all simulations.  

\begin{figure}[ht]
        \centering \includegraphics[width=1\linewidth]
        {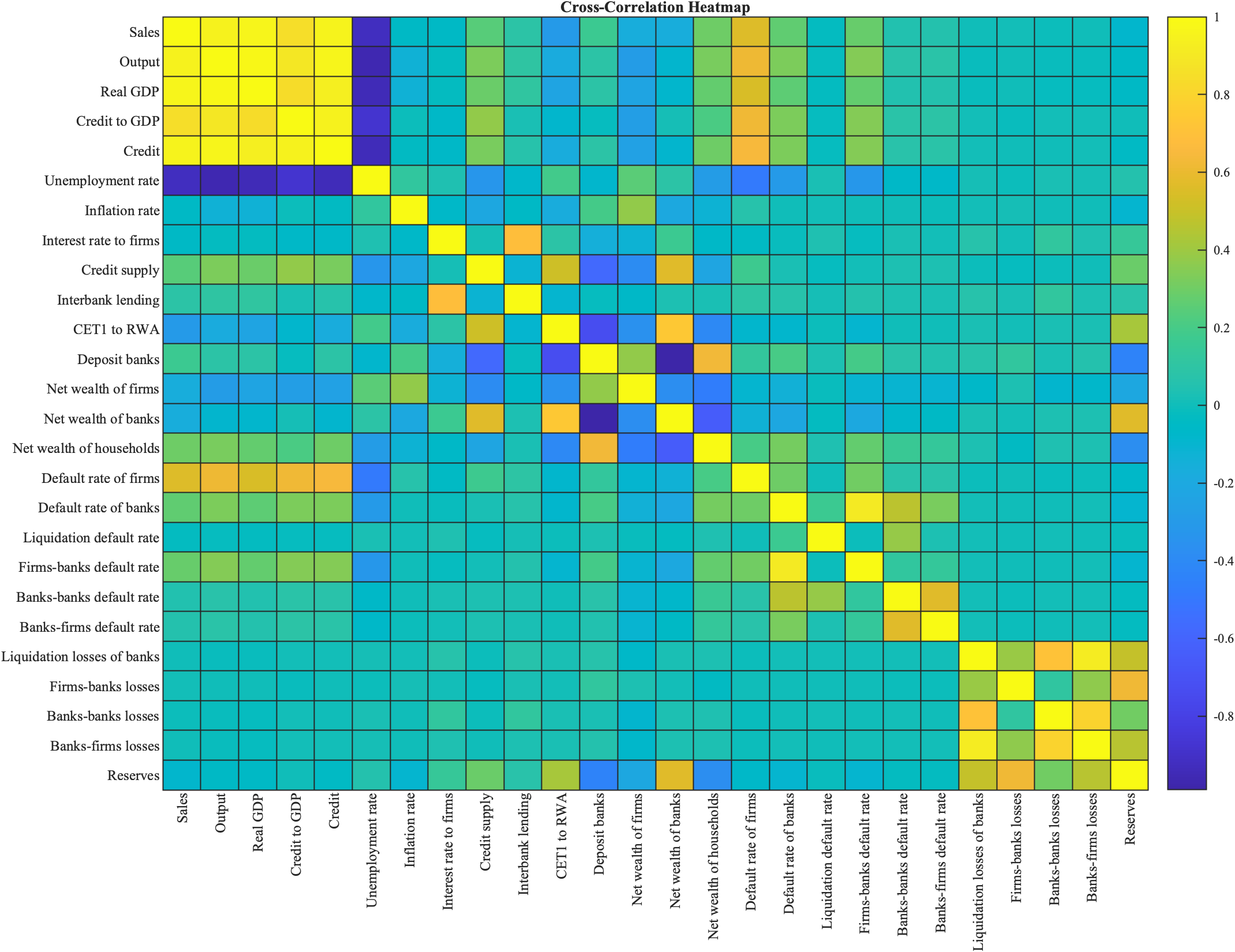}
        \caption{Cross-correlation between main variables for the baseline model.}
    \label{fig:crosscorr}
    \end{figure}

    In Table \ref{tab:scenario_comparison_baseline}, for the baseline scenario (no CBDC), we report the mean, median, standard deviation, $1\%$ and $99\%$ percentile ($P_{01}$ and $P_{99}$) for each variable. In Figure \ref{fig:crosscorr}, we report the cross-correlation heatmap highlighting the main relationships among the variables. We observe a strong positive correlation between output/real GDP and credit and sales. On average, the share of wealth retained as CBDC is $7.2, \ 41.2, \ 14.9, \ 17.6,\ \text{and}\ 19.5\%$ according to the five conversion rules.
    
    \FloatBarrier
    \FloatBarrier
    
\begin{figure}
        \centering
        \includegraphics{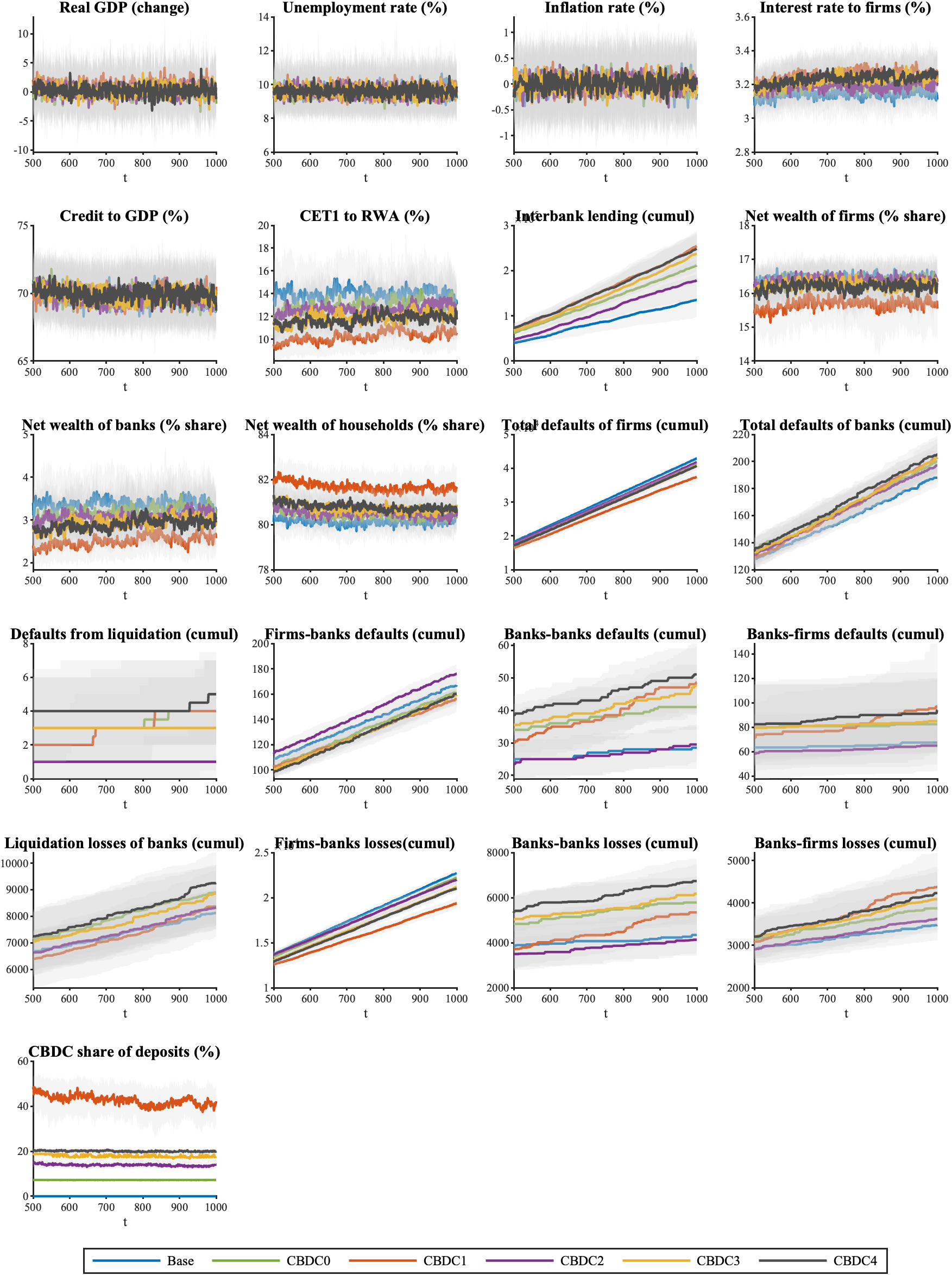}
        \caption{Median values for the time series. }
        \label{fig.cbdc_time_series}
    \end{figure}

\begin{figure}
        \centering
        \includegraphics
        {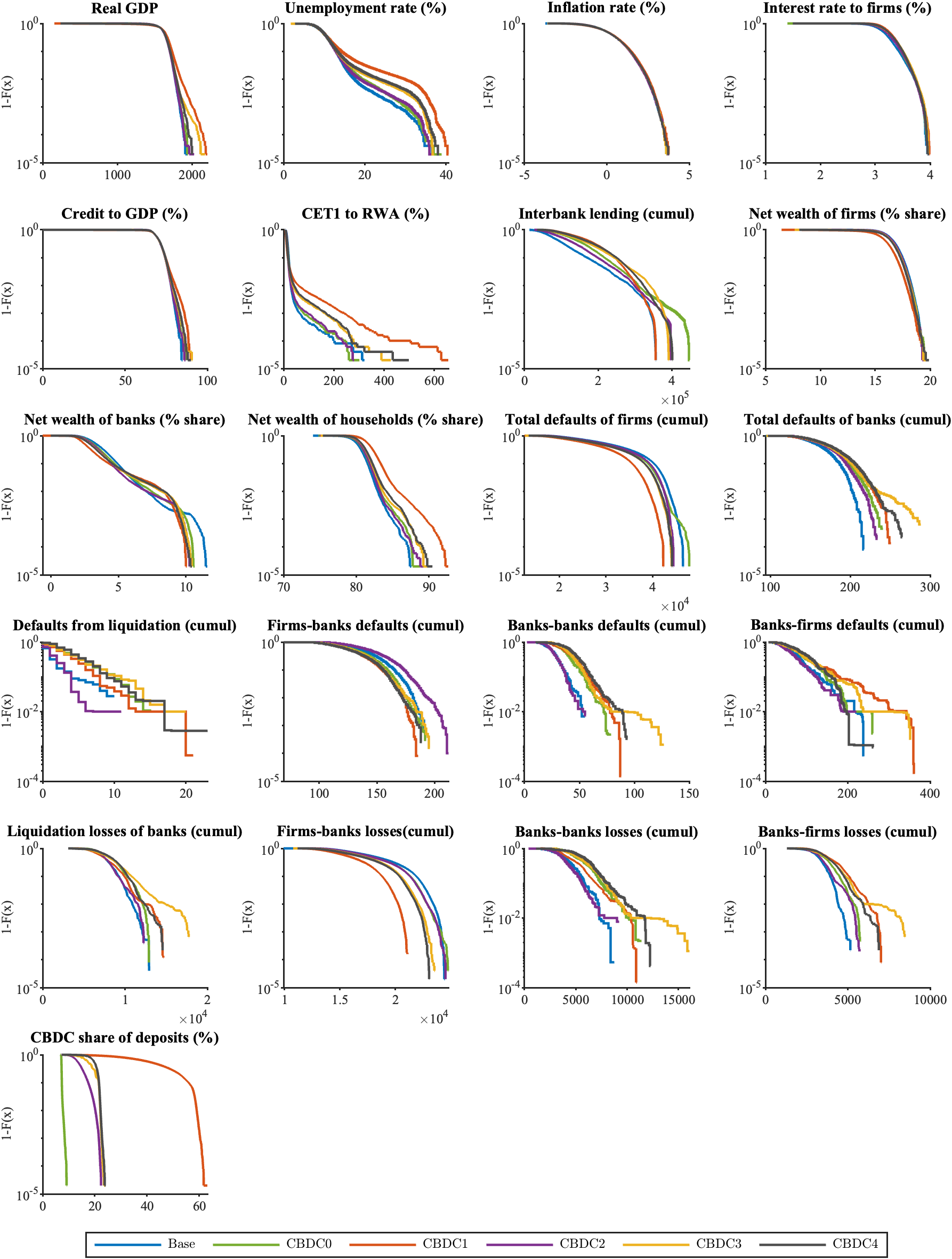}
        \caption{Complementary cumulative distribution functions (CCDFs) of selected variables, CCDF is defined as $1-F(x)$, where $F(x)$ is the cumulative distribution function of $x$.}
        \label{fig.cbdc_ccdf}
    \end{figure}

    Comparing the economy with CBDC to one without it, we develop the analysis in two directions through two sets of indicators. The first set of indicators includes mean, standard deviation, and median values, and the second the extreme values (percentiles 1\% and 99\%). The first set allows us to investigate the performance of the economy in normal times, and the other under extreme conditions.
    In Table \ref{tab:scenario_comparison_part1}-\ref{tab:scenario_comparison_part5}, we report the statistics for each variable considering the five rules determining the amount of CBDC.
    In Figure \ref{fig.cbdc_time_series}, we present the median values over time for the main variables considering the baseline scenario (no CBDC) and those with CBDC according to the different rules. 
    In Figure \ref{fig.cbdc_ccdf}, we report the complementary cumulative distribution functions of the same variables. 
    The figure describes how often, in the simulated data, a variable is above the value reported on the x-axis, providing a comprehensive description of the probability distribution of the different variables in the simulations.

    \FloatBarrier
    \begin{table}[ht]
\scriptsize
\centering
\setlength{\tabcolsep}{1pt}
\renewcommand{\arraystretch}{0.8}
\begin{tabular}{l|rrrrr}
\toprule
 & \multicolumn{5}{c}{\textbf{Base}} \\
 \textbf{Variable} & \textbf{mean} & \textbf{sd} & \textbf{med} & $\mathbf{P_{01}}$ & $\mathbf{P_{99}}$ \\
 \midrule
 Output & 1630.948 & 135.013 & 1641.005 & 1258.074 & 1895.049 \\
 Real GDP & 1596.738 & 107.540 & 1614.895 & 1258.074 & 1772.049 \\
 Unemployment rate (\%) & 9.832 & 2.597 & 9.560 & 5.240 & 17.440 \\
 Inflation rate (\%) & 0.002 & 1.071 & 0.006 & -2.419 & 2.442 \\
 Interest rate to firms (\%) & 3.149 & 0.218 & 3.142 & 2.645 & 3.713 \\
 Credit to GDP (\%) & 69.886 & 3.747 & 69.770 & 61.966 & 77.954 \\
 CET1 to RWA (\%) & 14.877 & 5.935 & 14.014 & 8.303 & 29.084 \\
 Interbank lending & 211.390 & 225.910 & 147.785 & 0.000 & 874.715 \\
 Net wealth of firms (\% share) & 16.352 & 0.975 & 16.424 & 13.433 & 18.304 \\
 Net wealth of banks (\% share) & 3.530 & 1.124 & 3.371 & 1.681 & 7.058 \\
 Net wealth of households (\% share) & 80.118 & 1.180 & 80.130 & 77.209 & 82.971 \\
 Default rate of firms (\%) & 9.831 & 7.301 & 9.600 & 0.000 & 29.200 \\
 Default rate of banks (\%) & 1.226 & 4.787 & 0.000 & 0.000 & 20.000 \\
 Liquidation default rate (\%) & 0.002 & 0.134 & 0.000 & 0.000 & 0.000 \\
 Firms-banks default rate (\%) & 1.211 & 4.394 & 0.000 & 0.000 & 20.000 \\
 Banks-banks default rate (\%) & 0.086 & 1.697 & 0.000 & 0.000 & 0.000 \\
 Banks-firms default rate (\%) & 0.002 & 0.038 & 0.000 & 0.000 & 0.000 \\
 Liquidation losses of banks to gdp (\%) & 0.351 & 0.081 & 0.341 & 0.196 & 0.578 \\
 Firms-banks losses to gdp (\%) & 0.846 & 0.062 & 0.839 & 0.708 & 1.000 \\
 Banks-banks losses to gdp (\%) & 0.196 & 0.077 & 0.187 & 0.063 & 0.424 \\
 Banks-firms losses to gdp (\%) & 0.151 & 0.036 & 0.147 & 0.080 & 0.248 \\
 \bottomrule
\end{tabular}
\caption{Summary statistics for baseline scenario. Mean (mean), standard deviation (sd), median (med), 1st ($P_{01}$) and 99th ($P_{99}$) percentile). }
\label{tab:scenario_comparison_baseline}
\end{table}

    \begin{table}[ht]
\scriptsize
\centering
\setlength{\tabcolsep}{1pt}
\renewcommand{\arraystretch}{0.8}
\begin{tabular}{l|rrrrrr}
 \toprule
 & \multicolumn{6}{c}{\textbf{CBDC0}} \\
 \textbf{Variable} & \textbf{dev} & \textbf{mean} & \textbf{sd} & \textbf{med} & $\mathbf{P_{01}}$ & $\mathbf{P_{99}}$ \\
 \midrule
 Output & -0.149 & 1628.524$^{***}$ & 144.095 & 1642.046 & 1151.578 & 1896.318 \\
 Real GDP & -0.099 & 1595.150$^{**\phantom{*}}$ & 118.797 & 1616.272 & 1151.578 & 1778.701 \\
 Unemployment rate (\%) & 0.083 & 9.914$^{***}$ & 2.786 & 9.600 & 5.240 & 19.360 \\
 Inflation rate (\%) & -0.000 & 0.002$^{\phantom{***}}$ & 1.078 & 0.011 & -2.454 & 2.445 \\
 Interest rate to firms (\%) & 0.057 & 3.206$^{***}$ & 0.204 & 3.200 & 2.738 & 3.720 \\
 Credit to GDP (\%) & 0.034 & 69.920$^{\phantom{***}}$ & 3.920 & 69.880 & 60.538 & 78.222 \\
 CET1 to RWA (\%) & -0.956 & 13.921$^{***}$ & 6.442 & 12.853 & 8.023 & 31.435 \\
 Interbank lending & 44.641 & 305.756$^{***}$ & 240.552 & 280.539 & 0.000 & 937.113 \\
 Net wealth of firms (\% share) & -0.121 & 16.231$^{***}$ & 1.059 & 16.334 & 12.601 & 18.246 \\
 Net wealth of banks (\% share) & -0.183 & 3.348$^{***}$ & 1.150 & 3.151 & 1.536 & 7.446 \\
 Net wealth of households (\% share) & 0.304 & 80.422$^{***}$ & 1.173 & 80.423 & 77.560 & 83.318 \\
 Default rate of firms (\%) & -0.156 & 9.675$^{***}$ & 7.267 & 9.400 & 0.000 & 29.000 \\
 Default rate of banks (\%) & 0.060 & 1.286$^{*\phantom{**}}$ & 5.103 & 0.000 & 0.000 & 30.000 \\
 Liquidation default rate (\%) & 0.004 & 0.006$^{***}$ & 0.264 & 0.000 & 0.000 & 0.000 \\
 Firms-banks default rate (\%) & -0.035 & 1.177$^{\phantom{***}}$ & 4.236 & 0.000 & 0.000 & 20.000 \\
 Banks-banks default rate (\%) & 0.071 & 0.156$^{***}$ & 2.307 & 0.000 & 0.000 & 0.000 \\
 Banks-firms default rate (\%) & 0.000 & 0.002$^{\phantom{***}}$ & 0.044 & 0.000 & 0.000 & 0.000 \\
 Liquidation losses of banks to gdp (\%) & 0.034 & 0.385$^{***}$ & 0.096 & 0.372 & 0.219 & 0.660 \\
 Firms-banks losses to gdp (\%) & -0.013 & 0.833$^{***}$ & 0.062 & 0.823 & 0.731 & 1.000 \\
 Banks-banks losses to gdp (\%) & 0.068 & 0.264$^{***}$ & 0.089 & 0.250 & 0.112 & 0.522 \\
 Banks-firms losses to gdp (\%) & 0.018 & 0.169$^{***}$ & 0.041 & 0.163 & 0.093 & 0.288 \\
 \bottomrule
 \multicolumn{7}{l}{\textit{Significance levels:} *** p<0.01, ** p<0.05, * p<0.1} \\
\end{tabular}
\caption{Comparison of scenarios: CBDC0 rule. Deviation from baseline mean (dev), mean (mean), standard deviation (sd), median (med), 1st ($P_{01}$) and 99th ($P_{99}$) percentile. The deviation from the baseline is expressed in percentage points for rates and ratios or as percentage change for variables in level (real GDP, interbank lending).}
\label{tab:scenario_comparison_part1}
\end{table}

    \begin{table}[ht]
\scriptsize
\centering
\setlength{\tabcolsep}{1pt}
\renewcommand{\arraystretch}{0.8}
\begin{tabular}{l|rrrrrr}
 \toprule
 & \multicolumn{6}{c}{\textbf{CBDC1}} \\
 \textbf{Variable} & \textbf{dev} & \textbf{mean} & \textbf{sd} & \textbf{med} & $\mathbf{P_{01}}$ & $\mathbf{P_{99}}$ \\
 \midrule
 Output & -0.905 & 1616.192$^{***}$ & 212.428 & 1653.007 & 643.733 & 1952.144 \\
 Real GDP & -0.517 & 1588.490$^{***}$ & 193.505 & 1632.796 & 643.733 & 1849.405 \\
 Unemployment rate (\%) & 0.587 & 10.419$^{***}$ & 4.108 & 9.600 & 4.720 & 29.760 \\
 Inflation rate (\%) & 0.002 & 0.003$^{\phantom{***}}$ & 1.156 & 0.031 & -2.564 & 2.549 \\
 Interest rate to firms (\%) & 0.113 & 3.262$^{***}$ & 0.186 & 3.256 & 2.844 & 3.721 \\
 Credit to GDP (\%) & -0.366 & 69.520$^{***}$ & 5.909 & 69.809 & 50.781 & 80.488 \\
 CET1 to RWA (\%) & -1.923 & 12.954$^{***}$ & 15.631 & 10.201 & 7.667 & 58.776 \\
 Interbank lending & 70.852 & 361.163$^{***}$ & 237.783 & 341.688 & 0.000 & 963.070 \\
 Net wealth of firms (\% share) & -0.968 & 15.384$^{***}$ & 1.578 & 15.670 & 9.171 & 17.832 \\
 Net wealth of banks (\% share) & -0.675 & 2.855$^{***}$ & 1.294 & 2.507 & 1.131 & 8.323 \\
 Net wealth of households (\% share) & 1.643 & 81.761$^{***}$ & 1.363 & 81.690 & 78.816 & 85.717 \\
 Default rate of firms (\%) & -1.350 & 8.480$^{***}$ & 7.462 & 7.800 & 0.000 & 29.600 \\
 Default rate of banks (\%) & 0.235 & 1.461$^{***}$ & 6.121 & 0.000 & 0.000 & 30.000 \\
 Liquidation default rate (\%) & 0.043 & 0.044$^{***}$ & 0.760 & 0.000 & 0.000 & 0.000 \\
 Firms-banks default rate (\%) & -0.140 & 1.072$^{***}$ & 3.993 & 0.000 & 0.000 & 20.000 \\
 Banks-banks default rate (\%) & 0.282 & 0.368$^{***}$ & 3.523 & 0.000 & 0.000 & 10.000 \\
 Banks-firms default rate (\%) & 0.010 & 0.012$^{***}$ & 0.222 & 0.000 & 0.000 & 0.200 \\
 Liquidation losses of banks to gdp (\%) & 0.007 & 0.358$^{***}$ & 0.094 & 0.342 & 0.213 & 0.638 \\
 Firms-banks losses to gdp (\%) & -0.095 & 0.751$^{***}$ & 0.072 & 0.739 & 0.624 & 0.941 \\
 Banks-banks losses to gdp (\%) & 0.031 & 0.227$^{***}$ & 0.088 & 0.215 & 0.090 & 0.529 \\
 Banks-firms losses to gdp (\%) & 0.032 & 0.183$^{***}$ & 0.045 & 0.178 & 0.111 & 0.316 \\
 \bottomrule
 \multicolumn{7}{l}{\textit{Significance levels:} *** p<0.01, ** p<0.05, * p<0.1} \\
\end{tabular}
\caption{Comparison of scenarios: CBDC1 rule. Deviation from baseline mean (dev), mean (mean), standard deviation (sd), median (med), 1st ($P_{01}$) and 99th ($P_{99}$) percentile. The deviation from the baseline is expressed in percentage points for rates and ratios or as percentage change for variables in level (real GDP, interbank lending).}
\label{tab:scenario_comparison_part2}
\end{table}

    \begin{table}[ht]
\scriptsize
\centering
\setlength{\tabcolsep}{1pt}
\renewcommand{\arraystretch}{0.8}
\begin{tabular}{l|rrrrrr}
 \toprule
 & \multicolumn{6}{c}{\textbf{CBDC2}} \\
 \textbf{Variable} & \textbf{dev} & \textbf{mean} & \textbf{sd} & \textbf{med} & $\mathbf{P_{01}}$ & $\mathbf{P_{99}}$ \\
 \midrule
 Output & -0.090 & 1629.480$^{*\phantom{**}}$ & 141.301 & 1642.179 & 1183.503 & 1891.873 \\
 Real GDP & -0.015 & 1596.497$^{\phantom{***}}$ & 116.113 & 1616.488 & 1183.503 & 1773.615 \\
 Unemployment rate (\%) & 0.065 & 9.897$^{***}$ & 2.739 & 9.600 & 5.280 & 18.800 \\
 Inflation rate (\%) & -0.002 & -0.001$^{\phantom{***}}$ & 1.068 & 0.004 & -2.411 & 2.433 \\
 Interest rate to firms (\%) & 0.046 & 3.195$^{***}$ & 0.212 & 3.187 & 2.713 & 3.722 \\
 Credit to GDP (\%) & -0.036 & 69.850$^{\phantom{***}}$ & 3.945 & 69.803 & 61.293 & 78.018 \\
 CET1 to RWA (\%) & -1.310 & 13.567$^{***}$ & 6.549 & 12.590 & 8.016 & 30.054 \\
 Interbank lending & 31.021 & 276.965$^{***}$ & 239.211 & 240.123 & 0.000 & 901.990 \\
 Net wealth of firms (\% share) & -0.086 & 16.266$^{***}$ & 1.023 & 16.358 & 12.912 & 18.219 \\
 Net wealth of banks (\% share) & -0.296 & 3.234$^{***}$ & 1.071 & 3.066 & 1.560 & 7.116 \\
 Net wealth of households (\% share) & 0.383 & 80.500$^{***}$ & 1.123 & 80.497 & 77.748 & 83.239 \\
 Default rate of firms (\%) & -0.166 & 9.665$^{***}$ & 7.254 & 9.400 & 0.000 & 29.000 \\
 Default rate of banks (\%) & 0.127 & 1.353$^{***}$ & 5.107 & 0.000 & 0.000 & 20.000 \\
 Liquidation default rate (\%) & 0.003 & 0.004$^{**\phantom{*}}$ & 0.210 & 0.000 & 0.000 & 0.000 \\
 Firms-banks default rate (\%) & 0.080 & 1.291$^{***}$ & 4.500 & 0.000 & 0.000 & 20.000 \\
 Banks-banks default rate (\%) & 0.033 & 0.119$^{***}$ & 2.029 & 0.000 & 0.000 & 0.000 \\
 Banks-firms default rate (\%) & 0.001 & 0.002$^{**\phantom{*}}$ & 0.050 & 0.000 & 0.000 & 0.000 \\
 Liquidation losses of banks to gdp (\%) & 0.002 & 0.353$^{***}$ & 0.080 & 0.343 & 0.212 & 0.587 \\
 Firms-banks losses to gdp (\%) & -0.016 & 0.830$^{***}$ & 0.063 & 0.820 & 0.722 & 0.994 \\
 Banks-banks losses to gdp (\%) & -0.009 & 0.187$^{***}$ & 0.071 & 0.177 & 0.045 & 0.407 \\
 Banks-firms losses to gdp (\%) & 0.006 & 0.157$^{***}$ & 0.038 & 0.151 & 0.093 & 0.272 \\
 \bottomrule
 \multicolumn{7}{l}{\textit{Significance levels:} *** p<0.01, ** p<0.05, * p<0.1} \\
\end{tabular}
\caption{Comparison of scenarios: CBDC2 rule. Deviation from baseline mean (dev), mean (mean), standard deviation (sd), median (med), 1st ($P_{01}$) and 99th ($P_{99}$) percentile. The deviation from the baseline is expressed in percentage points for rates and ratios or as percentage change for variables in level (real GDP, interbank lending).}
\label{tab:scenario_comparison_part3}
\end{table}

    \begin{table}[ht]
\scriptsize
\centering
\setlength{\tabcolsep}{1pt}
\renewcommand{\arraystretch}{0.8}
\begin{tabular}{l|rrrrrr}
 \toprule
 & \multicolumn{6}{c}{\textbf{CBDC3}} \\
 \textbf{Variable} & \textbf{dev} & \textbf{mean} & \textbf{sd} & \textbf{med} & $\mathbf{P_{01}}$ & $\mathbf{P_{99}}$ \\
 \midrule
 Output & -0.363 & 1625.032$^{***}$ & 163.756 & 1644.434 & 980.881 & 1908.625 \\
 Real GDP & -0.237 & 1592.957$^{***}$ & 141.114 & 1619.662 & 980.881 & 1793.195 \\
 Unemployment rate (\%) & 0.205 & 10.037$^{***}$ & 3.173 & 9.560 & 5.120 & 22.840 \\
 Inflation rate (\%) & -0.003 & -0.001$^{\phantom{***}}$ & 1.096 & 0.009 & -2.481 & 2.482 \\
 Interest rate to firms (\%) & 0.088 & 3.236$^{***}$ & 0.208 & 3.228 & 2.774 & 3.747 \\
 Credit to GDP (\%) & -0.104 & 69.782$^{***}$ & 4.658 & 69.900 & 58.324 & 78.635 \\
 CET1 to RWA (\%) & -1.421 & 13.456$^{***}$ & 9.682 & 11.840 & 7.855 & 37.878 \\
 Interbank lending & 65.286 & 349.398$^{***}$ & 256.957 & 321.721 & 0.000 & 993.164 \\
 Net wealth of firms (\% share) & -0.262 & 16.090$^{***}$ & 1.185 & 16.236 & 11.424 & 18.173 \\
 Net wealth of banks (\% share) & -0.353 & 3.178$^{***}$ & 1.225 & 2.925 & 1.448 & 8.175 \\
 Net wealth of households (\% share) & 0.614 & 80.732$^{***}$ & 1.183 & 80.731 & 77.881 & 83.686 \\
 Default rate of firms (\%) & -0.377 & 9.454$^{***}$ & 7.312 & 9.200 & 0.000 & 29.200 \\
 Default rate of banks (\%) & 0.139 & 1.365$^{***}$ & 5.456 & 0.000 & 0.000 & 30.000 \\
 Liquidation default rate (\%) & 0.011 & 0.013$^{***}$ & 0.384 & 0.000 & 0.000 & 0.000 \\
 Firms-banks default rate (\%) & -0.037 & 1.175$^{\phantom{***}}$ & 4.175 & 0.000 & 0.000 & 20.000 \\
 Banks-banks default rate (\%) & 0.140 & 0.226$^{***}$ & 2.816 & 0.000 & 0.000 & 0.000 \\
 Banks-firms default rate (\%) & 0.001 & 0.003$^{***}$ & 0.063 & 0.000 & 0.000 & 0.000 \\
 Liquidation losses of banks to gdp (\%) & 0.035 & 0.386$^{***}$ & 0.118 & 0.365 & 0.195 & 0.760 \\
 Firms-banks losses to gdp (\%) & -0.041 & 0.804$^{***}$ & 0.065 & 0.794 & 0.693 & 0.982 \\
 Banks-banks losses to gdp (\%) & 0.076 & 0.272$^{***}$ & 0.101 & 0.257 & 0.092 & 0.600 \\
 Banks-firms losses to gdp (\%) & 0.027 & 0.177$^{***}$ & 0.051 & 0.169 & 0.090 & 0.330 \\
 \bottomrule
 \multicolumn{7}{l}{\textit{Significance levels:} *** p<0.01, ** p<0.05, * p<0.1} \\
\end{tabular}
\caption{Comparison of scenarios: CBDC3 rule. Deviation from baseline mean (dev), mean (mean), standard deviation (sd), median (med), 1st ($P_{01}$) and 99th ($P_{99}$) percentile. The deviation from the baseline is expressed in percentage points for rates and ratios or as percentage change for variables in level (real GDP, interbank lending).}
\label{tab:scenario_comparison_part4}
\end{table}

    \begin{table}[ht]
\scriptsize
\centering
\setlength{\tabcolsep}{1pt}
\renewcommand{\arraystretch}{0.8}
\begin{tabular}{l|rrrrrr}
 \toprule
 & \multicolumn{6}{c}{\textbf{CBDC4}} \\
 \textbf{Variable} & \textbf{dev} & \textbf{mean} & \textbf{sd} & \textbf{med} & $\mathbf{P_{01}}$ & $\mathbf{P_{99}}$ \\
 \midrule
 Output & -0.484 & 1623.055$^{***}$ & 168.617 & 1644.334 & 940.086 & 1904.701 \\
 Real GDP & -0.350 & 1591.144$^{***}$ & 146.483 & 1619.531 & 940.086 & 1791.149 \\
 Unemployment rate (\%) & 0.258 & 10.090$^{***}$ & 3.277 & 9.600 & 5.240 & 23.660 \\
 Inflation rate (\%) & 0.001 & 0.003$^{\phantom{***}}$ & 1.104 & 0.011 & -2.477 & 2.489 \\
 Interest rate to firms (\%) & 0.091 & 3.240$^{***}$ & 0.201 & 3.232 & 2.779 & 3.729 \\
 Credit to GDP (\%) & -0.057 & 69.829$^{**\phantom{*}}$ & 4.794 & 69.948 & 57.865 & 78.864 \\
 CET1 to RWA (\%) & -1.393 & 13.484$^{***}$ & 10.340 & 11.737 & 7.848 & 41.158 \\
 Interbank lending & 64.781 & 348.329$^{***}$ & 247.554 & 328.603 & 0.000 & 973.072 \\
 Net wealth of firms (\% share) & -0.337 & 16.014$^{***}$ & 1.243 & 16.194 & 11.076 & 18.127 \\
 Net wealth of banks (\% share) & -0.365 & 3.165$^{***}$ & 1.244 & 2.896 & 1.452 & 8.397 \\
 Net wealth of households (\% share) & 0.702 & 80.820$^{***}$ & 1.198 & 80.790 & 77.992 & 84.177 \\
 Default rate of firms (\%) & -0.437 & 9.394$^{***}$ & 7.330 & 9.000 & 0.000 & 29.700 \\
 Default rate of banks (\%) & 0.213 & 1.439$^{***}$ & 5.719 & 0.000 & 0.000 & 30.000 \\
 Liquidation default rate (\%) & 0.017 & 0.019$^{***}$ & 0.472 & 0.000 & 0.000 & 0.000 \\
 Firms-banks default rate (\%) & 0.002 & 1.214$^{\phantom{***}}$ & 4.300 & 0.000 & 0.000 & 20.000 \\
 Banks-banks default rate (\%) & 0.169 & 0.255$^{***}$ & 2.995 & 0.000 & 0.000 & 10.000 \\
 Banks-firms default rate (\%) & 0.002 & 0.004$^{***}$ & 0.100 & 0.000 & 0.000 & 0.000 \\
 Liquidation losses of banks to gdp (\%) & 0.036 & 0.387$^{***}$ & 0.098 & 0.380 & 0.185 & 0.652 \\
 Firms-banks losses to gdp (\%) & -0.048 & 0.798$^{***}$ & 0.062 & 0.788 & 0.686 & 0.961 \\
 Banks-banks losses to gdp (\%) & 0.087 & 0.283$^{***}$ & 0.098 & 0.280 & 0.080 & 0.580 \\
 Banks-firms losses to gdp (\%) & 0.027 & 0.177$^{***}$ & 0.044 & 0.175 & 0.081 & 0.297 \\
 \bottomrule
 \multicolumn{7}{l}{\textit{Significance levels:} *** p<0.01, ** p<0.05, * p<0.1} \\
\end{tabular}
\caption{Comparison of scenarios: CBDC4 rule. Deviation from baseline mean (dev), mean (mean), standard deviation (sd), median (med), 1st ($P_{01}$) and 99th ($P_{99}$) percentile. The deviation from the baseline is expressed in percentage points for rates and ratios or as percentage change for variables in level (real GDP, interbank lending).}
\label{tab:scenario_comparison_part5}
\end{table}

    \FloatBarrier

    Assuming a flat 10\% conversion rule of deposits in CBDC (CBDC0 rule), we observe an average negative effect on output and real GDP. Unemployment goes up but in a limited way. Although the differences with respect to the baseline model are small, they turn out to be statistically significant. 
    A different outcome is observed with the CBDC1 rule that allows for the substitution up to $80\%$ of deposits.
    On average, the reduction of GDP and the increase in unemployment are around $0.5\%$. 
    By construction the amount of money is fixed, and therefore almost no effect is observed on inflation in both scenarios. 

    In extreme events, limited effects are observed for a flat scenario (CBDC0 rule). 
    Instead, the CBDC1 rule produces significant effects: GDP could reduce to $51\%$ of the baseline scenario (1st percentile) and unemployment could go up to $29.7\%$ (from $17.4\%$) (99th percentile). 
    The results show that allowing for large withdrawals of deposits may exacerbate fluctuations in the economy.
    A tight cap on CBDC adoption and a deposit insurance scheme (CBDC2-4 rules) produce macroeconomic effects only marginally stronger than those observed under the CBDC0 rule. 

    The CBDC adoption affects the credit intermediation of banks. In all scenarios, the average interest rate on loans to firms goes up but for a limited amount (up to 11 basis points). Also, the credit in the economy on average displays a limited reduction (non significant in case of CBDC0 rule). It is interesting to notice that even in the high-substitution scenario (CBDC1 rule) the average effects are limited, half the effect estimated for the US in \citet{NYFFE}.

    CBDC affects the balance sheet of banks. 
    As suggested in \citet{ADAL}, in all scenarios, banks rely on the interbank market to meet liquidity needs due to substitution of deposits with CBDC. 
    The effect turns out to be the strongest under the CBDC1 rule both on average and in extreme events.   
    The substitution of deposits with CBDC renders banks less capitalized  (CET1 ratio decreases on average) with an increase in bank defaults. 
    This leads to more frequent liquidation of assets with larger losses for banks.  
    The phenomena are more pronounced in the large substitution scenario (CBDC1 rule). 

    Substitution of deposits into CBDC induces a redistribution of wealth, from firms and banks to households. 
    The redistribution is a consequence of the increased financial instability associated with the introduction of CBDC. Bank defaults triggered by the substitution of deposits with CBDC have asymmetric effects on banks, firms and households. 
    By diverting wealth to CBDC households are less vulnerable when banks default which prevents losses on bank deposits. As suggested in \citet{WILL2}, bank defaults are more frequent with CBDC but they hurt households less. 
    Instead, banks' wealth decreases and defaults go up, while firms' deposits are more exposed to second-round effects of financial contagion (banks-firms defaults and banks-firms losses to GDP go up). 

    The introduction of a CBDC tends to lower the default rate among firms. This result can be explained by the higher funding costs faced by banks, which, due to increased reliance on interbank borrowing, are passed on to firms through higher lending rates. As a consequence, firms experience lower profits and a tighter supply of credit, which limits their ability to fully implement production plans. While this leads to reduced output, it also decreases the likelihood of default. In practice, contagion from banks plays only a minor role in firm failures; most defaults originate from systematic errors in firms’ price and quantity adjustments. When credit becomes more constrained, such mistakes are less likely to result in insolvency, since firms depend more on their own resources to finance production. These dynamics are especially evident under the CBDC1 rule: the possibility of replacing a significant share of deposits with CBDC reduces firms’ net wealth and growth but also lowers their default rates. Overall, in a CBDC environment, firms tend to be smaller but financially sounder.

\begin{figure}
        \centering
        \includegraphics
        {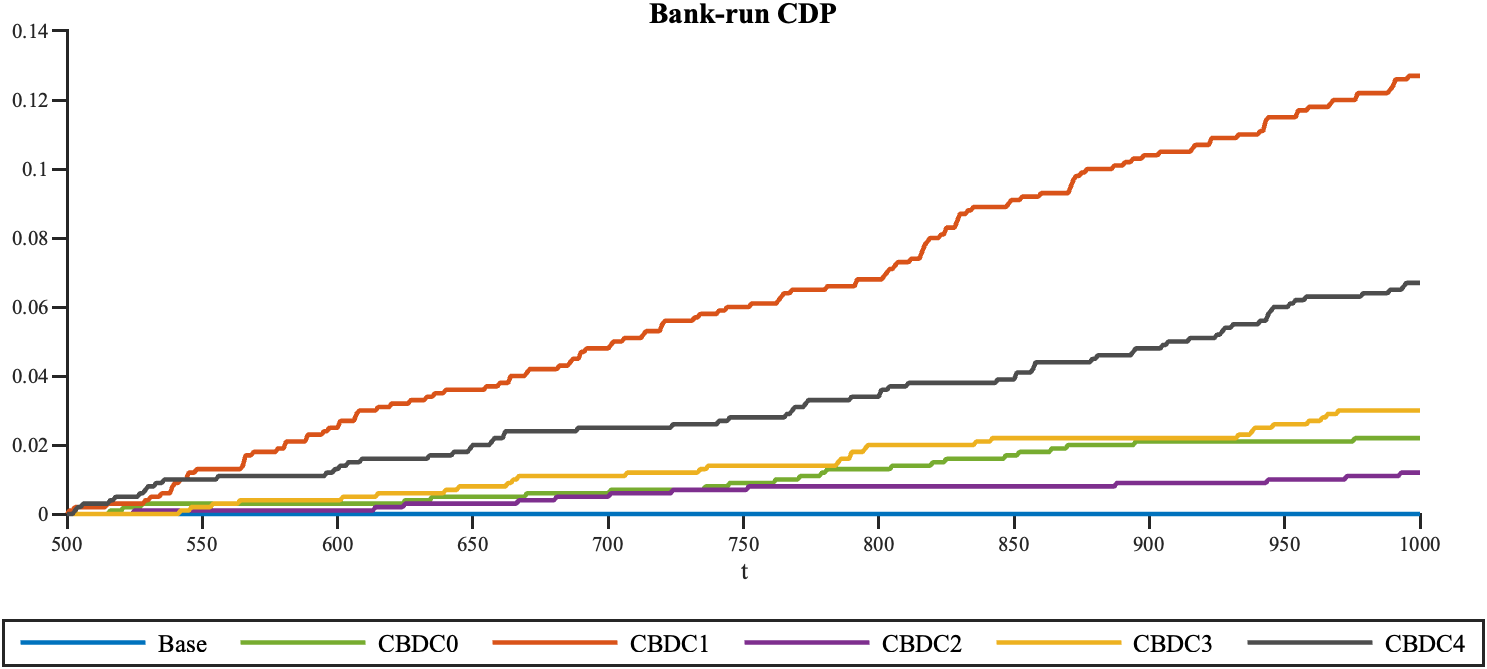}
        \caption{Average bank-run CDP of banks, log scale. The CDP represents the average probability (across banks and simulations) that a bank defaults by time $t$.}
        \label{fig.bank_run1-5}
    \end{figure}

    The conversion of deposits into CBDC plays a relevant role in driving bank defaults. 
    In Figure \ref{fig.bank_run1-5}, we report the average probability across banks and simulations of bank's default ignited by the conversion of deposits into CBDC (Cumulative Default Probability, CDP).
    We define this specific type of bank default as a \emph{"bank-run"}. A bank-run is characterized by the occurrence of the following events: (i) decline in bank's deposits due to conversion into CBDC;
    (ii) the bank becomes illiquid ($LR < 1$) and cannot borrow funds on the interbank market; 
    (iii) the bank's net wealth becomes negative as the result of asset liquidation.
    The probability of bank-run defaults is below $7\%$ for the conversion rules CBDC0, CBDC2-CBDC4. However, with a loose bound on the substitution of deposits (CBDC1 rule), the average cumulative probability of default of a bank after 1,000 iterations is roughly $12\%$. 
    It seems that, as the share of deposits that can be converted in CBDC increases, the interbank market becomes unable to allocate liquidity among banks, mostly because shortages are more severe and the liquidity in the system is  kept constant.
    Allowing for a high upper-bound on the conversion of deposits into CBDC, a flight-to-quality bank-run is more frequent leading to financial instability.

    Contagion in the financial system occurs through four channels. We assess their relevance under the different scenarios.

    \begin{enumerate}
        \item \emph{Losses (defaults) from liquidation} occur when the loss incurred by a bank selling its assets below their face values reduces the net wealth of the bank, eventually to a negative level which causes default.
        In the baseline model, CBDC0 and CBDC2-CBDC4 rules, both losses and defaults from banks' liquidation are limited on average and in the extreme events.
        In contrast, the possibility of a significant withdrawal rate of deposits (CBDC1 rule) leads to an increase both of banks' losses and defaults.
        \item \emph{Firms-banks losses (defaults)} occur when the default of a firm on loans reduces the net wealth of the bank. The different models (baseline, CBDC0-CBDC4 rule) show similar results (on average and in extreme events). 
        \item \emph{Banks-banks, or interbank, losses (defaults)} occur when a bank's counterparty is unable to repay its interbank obligations due to insolvency, thus reducing the net wealth of the lender banks. This channel presents similar results in the  baseline model, CBDC0, CBDC2-CBDC4 rules. The magnitude is much stronger in the case when the CBDC1 rule is applied.
        \item \emph{Banks-firms losses (defaults)} occur when the default of a bank leads to the default of one or more firms. 
        When the bank's net wealth becomes negative, it becomes insolvent and therefore may default on its deposits, thus reducing the net wealth of depositor firms. 
        This channel is stronger, although to a limited extent, in the scenarios with adoption of CBDC. 
    \end{enumerate}

\begin{figure}[H]
        \centering
        \includegraphics[width=1\linewidth]{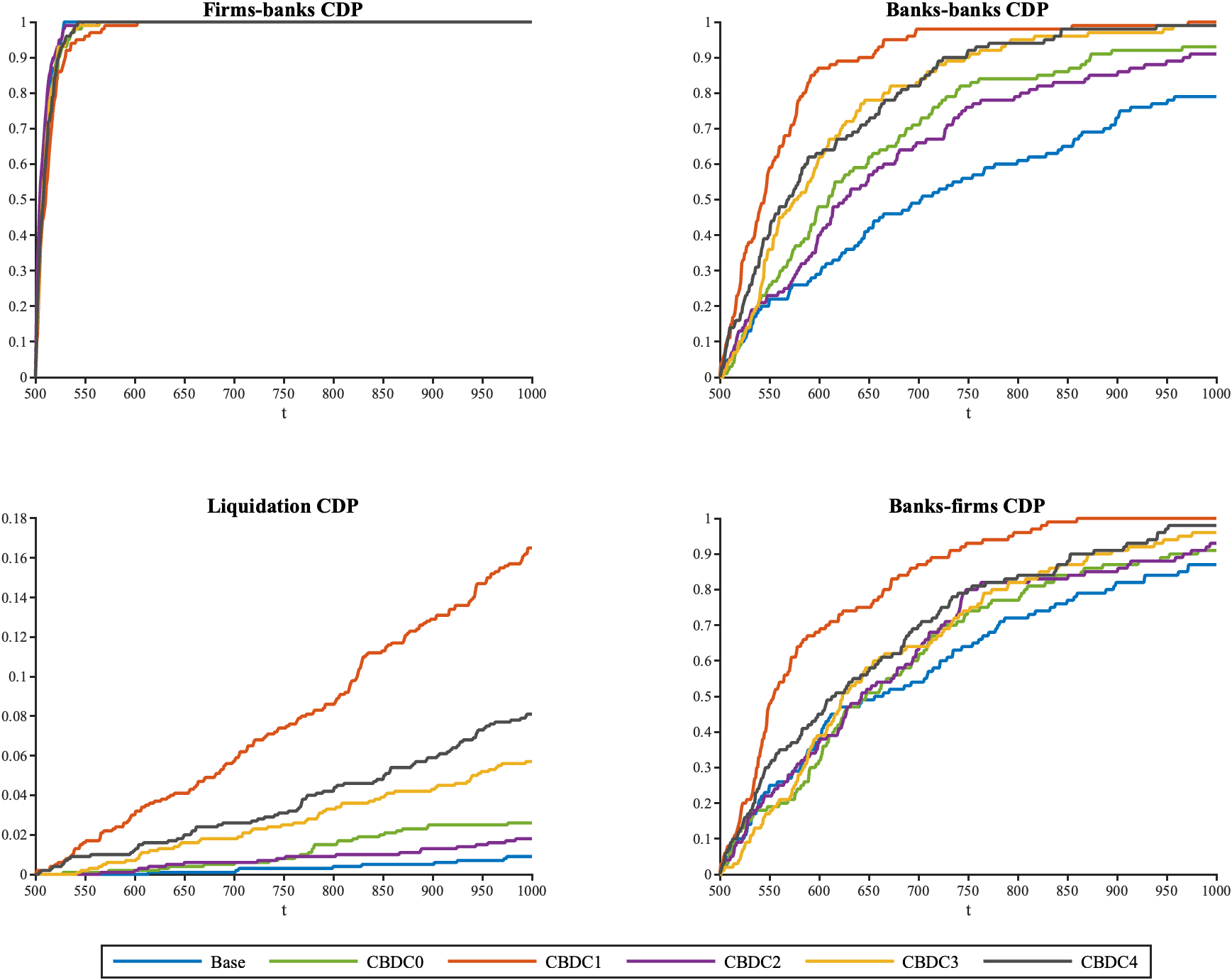}
        \caption{Average CDP of banks by type of default (firms-banks, banks-banks, liquidation, banks-firms). The (CDP) represents the average probability (across banks and simulations) that a bank defaults by time $t$.}
        \label{fig.CDP}
    \end{figure}

    In Figure \ref{fig.CDP}, we provide a representation of the CDP of a bank for the four channels. We confirm the above results. The firms-banks channel, which relates the firm default to the bank balance sheet, is by far the most relevant in all scenarios. The second and fourth channels (banks-banks and banks-firms) play a significant role in all scenarios.
    A more relevant effect is observed in the case when there is a large substitution of deposits with CBDC. The channel associated with the liquidation of assets plays a small role in the baseline scenario and a limited one in those with a low upper-bound conversion rate of deposits.  When a large share of withdrawal is allowed, the effect turns out to be much more relevant.
    This evidence confirms that the adoption of CBDC by households with few restrictions can cause financial instability.
    
    It is interesting to notice that financial instability implications are not only affected by the amount of CBDC but also by the smoothness of the conversion mechanism. 
    Under the CBDC2 scenario, the average share of wealth held in CBDC is 14.9\%, compared to 7.2\% under CBDC0. Nevertheless, financial instability is less pronounced in the CBDC2 scenario than in CBDC0. Figures \ref{fig.cbdc_time_series} and \ref{fig.cbdc_ccdf} show that defaults of banks driven by two key channels (liquidation of assets and interbank defaults) with the CBDC2 rule are the closest to those observed in the baseline scenario. The observation is confirmed  in Figure \ref{fig.bank_run1-5} for bank-run CDP, and in Figure \ref{fig.CDP} for both banks-banks and liquidation CDP. For ease of comparison, Table \ref{tab:means_by_scenario} reports the mean values across scenarios.
    These findings suggest that, from a macro-financial perspective, a smooth conversion mechanism dependent on the perceived riskiness of banks (CBDC2 rule) is less unstable than a flat rule (CBDC0 rule) or a 0-1 choice (CBDC3 and CBDC4 rule) dependent on the riskiness of banks even though it leads to the withdraw, on average, of more liquidity from the banking sector. 
    This observation leads us to investigate the social welfare implications and the optimal quantity of CBDC in Section \ref{sec.soc}.

    \begin{table}[h]
\scriptsize
\centering
\setlength{\tabcolsep}{3pt}
\renewcommand{\arraystretch}{0.9}
\begin{tabular}{l|rrrrrr}
\toprule
 & \textbf{Base} & \textbf{CBDC0} & \textbf{CBDC1} & \textbf{CBDC2} & \textbf{CBDC3} & \textbf{CBDC4} \\
\midrule
Output & 1630.948 & 1628.524$^{***}$ & 1616.192$^{***}$ & 1629.480$^{*\phantom{**}}$ & 1625.032$^{***}$ & 1623.055$^{***}$ \\
Real GDP & 1596.738 & 1595.150$^{**\phantom{*}}$ & 1588.490$^{***}$ & 1596.497$^{\phantom{***}}$ & 1592.957$^{***}$ & 1591.144$^{***}$ \\
Unemployment rate (\%) & 9.832 & 9.914$^{***}$ & 10.419$^{***}$ & 9.897$^{***}$ & 10.037$^{***}$ & 10.090$^{***}$ \\
Inflation rate (\%) & 0.002 & 0.002$^{\phantom{***}}$ & 0.003$^{\phantom{***}}$ & -0.001$^{\phantom{***}}$ & -0.001$^{\phantom{***}}$ & 0.003$^{\phantom{***}}$ \\
Interest rate to firms (\%) & 3.149 & 3.206$^{***}$ & 3.262$^{***}$ & 3.195$^{***}$ & 3.236$^{***}$ & 3.240$^{***}$ \\
Credit to GDP (\%) & 69.886 & 69.920$^{\phantom{***}}$ & 69.520$^{***}$ & 69.850$^{\phantom{***}}$ & 69.782$^{***}$ & 69.829$^{**\phantom{*}}$ \\
CET1 to RWA (\%) & 14.877 & 13.921$^{***}$ & 12.954$^{***}$ & 13.567$^{***}$ & 13.456$^{***}$ & 13.484$^{***}$ \\
Interbank lending & 211.390 & 305.756$^{***}$ & 361.163$^{***}$ & 276.965$^{***}$ & 349.398$^{***}$ & 348.329$^{***}$ \\
Net wealth of firms (\% share) & 16.352 & 16.231$^{***}$ & 15.384$^{***}$ & 16.266$^{***}$ & 16.090$^{***}$ & 16.014$^{***}$ \\
Net wealth of banks (\% share) & 3.530 & 3.348$^{***}$ & 2.855$^{***}$ & 3.234$^{***}$ & 3.178$^{***}$ & 3.165$^{***}$ \\
Net wealth of households (\% share) & 80.118 & 80.422$^{***}$ & 81.761$^{***}$ & 80.500$^{***}$ & 80.732$^{***}$ & 80.820$^{***}$ \\
Default rate of firms (\%) & 9.831 & 9.675$^{***}$ & 8.480$^{***}$ & 9.665$^{***}$ & 9.454$^{***}$ & 9.394$^{***}$ \\
Default rate of banks (\%) & 1.226 & 1.286$^{*\phantom{**}}$ & 1.461$^{***}$ & 1.353$^{***}$ & 1.365$^{***}$ & 1.439$^{***}$ \\
Liquidation default rate (\%) & 0.002 & 0.006$^{***}$ & 0.044$^{***}$ & 0.004$^{**\phantom{*}}$ & 0.013$^{***}$ & 0.019$^{***}$ \\
Firms-banks default rate (\%) & 1.211 & 1.177$^{\phantom{***}}$ & 1.072$^{***}$ & 1.291$^{***}$ & 1.175$^{\phantom{***}}$ & 1.214$^{\phantom{***}}$ \\
Banks-banks default rate (\%) & 0.086 & 0.156$^{***}$ & 0.368$^{***}$ & 0.119$^{***}$ & 0.226$^{***}$ & 0.255$^{***}$ \\
Banks-firms default rate (\%) & 0.002 & 0.002$^{\phantom{***}}$ & 0.012$^{***}$ & 0.002$^{**\phantom{*}}$ & 0.003$^{***}$ & 0.004$^{***}$ \\
Liquidation losses of banks to gdp (\%) & 0.351 & 0.385$^{***}$ & 0.358$^{***}$ & 0.353$^{***}$ & 0.386$^{***}$ & 0.387$^{***}$ \\
Firms-banks losses to gdp (\%) & 0.846 & 0.833$^{***}$ & 0.751$^{***}$ & 0.830$^{***}$ & 0.804$^{***}$ & 0.798$^{***}$ \\
Banks-banks losses to gdp (\%) & 0.196 & 0.264$^{***}$ & 0.227$^{***}$ & 0.187$^{***}$ & 0.272$^{***}$ & 0.283$^{***}$ \\
Banks-firms losses to gdp (\%) & 0.151 & 0.169$^{***}$ & 0.183$^{***}$ & 0.157$^{***}$ & 0.177$^{***}$ & 0.177$^{***}$ \\
\bottomrule
 \multicolumn{6}{l}{\textit{Significance levels:} *** p<0.01, ** p<0.05, * p<0.1} \\
\end{tabular}
\caption{Mean values by scenario with statistical significance compared to Base.}
\label{tab:means_by_scenario}
\end{table}

    \section{Social Welfare Evaluation}
    \label{sec.soc}

    The adoption of CBDC by households leads to a redistribution of wealth from firms and banks to households.
    We observe an increase in the quota of wealth retained by households mostly because defaults of banks go up, firms are strongly affected by banks' defaults and households are less affected as they retain part of their wealth as CBDC. 

    We investigate the welfare implications of CBDC adoption by evaluating the distribution of wealth through a social welfare function. 
    We consider two different social welfare functions, namely the Atkinson ($\text{SWF}^{atk}$) function \citep{atkinson1970measurement} and the mean-variance ($\text{SWF}^{mv}$) function.
    Both functions return a score that takes into account relative wealth, compared to the average level, and its dispersion. The highest social welfare is achieved when the score is equal to one.

    The Atkinson social welfare function is
    \begin{equation}\label{eq.SWF_atkinson}
    \text{SWF}^{atk} =  
    \begin{cases}
        \left[ \frac{1}{N}\sum_{i=1}^{N} \left(\frac{x_i}{\bar{x}}\right)^{1-    \varepsilon} \right]^{\tfrac{1}{1-\varepsilon}} & \text{if }     \varepsilon \neq 1 \\[1em]
        \prod_{i=1}^{N} \left(\frac{x_i}{\bar{x}}\right)^{1/N} & \text{if } \varepsilon = 1\;\;,
    \end{cases}
    \end{equation}
    where $N$ is the number of observations, $\bar{x}$ is the mean value of the observations, and the parameter $\varepsilon$ represents the aversion to inequality (higher values give more weight to lower wealth).

    The mean-variance social welfare function is the arithmetic mean across households minus the variance multiplied by a parameter $\lambda$ that captures inequality aversion.  
    \begin{equation}\label{eq.SWF_meanvariance}    
        \text{SWF}^{mv} = \frac{1}{N}\sum_{i=1}^{N} \left(\frac{x_i}{\bar{x}}\right) -  \frac{\lambda}{N}\sum_{i=1}^{N} \left(\frac{x_i - \bar{x}}{\bar{x}} \right)^2 
        = 1 - \frac{\lambda}{N}\sum_{i=1}^{N} \left(\frac{x_i}{\bar{x}} -1\right)^2\;\;.
    \end{equation}

    Table \ref{NW} reports the social welfare scores evaluated on net wealth of households for the baseline scenario and the five CBDC rules. 
    The columns show social welfare scores for different inequality aversion parameters; the parameters for the least inequality-averse society are $\varepsilon=0.5$ and $\lambda=0.25$, while those for the most inequality-averse society are $\varepsilon=2$ and $\lambda=1$. Notice that the social welfare function levels are quite similar because the model presents very little heterogeneity (consumers share the same productivity, wages, government transfers, and initial wealth).

    \begin{table}[H]
    \scriptsize
    \centering
    \begin{tabular}{l|llll|llll}
\toprule
Scenario & \multicolumn{4}{c|}{Atkinson} & \multicolumn{4}{c}{Mean-variance} \\
 & $\varepsilon=0.5$ & $\varepsilon=1$ & $\varepsilon=1.5$ & $\varepsilon=2$ & $\lambda=0.25$ & $\lambda=0.5$ & $\lambda=0.75$ & $\lambda=1$ \\
\midrule
Base  & 0.9937$^{\phantom{***}}$ & 0.9871$^{\phantom{***}}$ & 0.9797$^{\phantom{***}}$ & 0.9666$^{\phantom{***}}$ & 0.9938$^{\phantom{***}}$ & 0.9877$^{\phantom{***}}$ & 0.9815$^{\phantom{***}}$ & 0.9753$^{\phantom{***}}$ \\

CBDC0 & 0.9938$^{**\phantom{*}}$ & 0.9873$^{**\phantom{*}}$ & 0.9801$^{**\phantom{*}}$ & 0.9666$^{\phantom{***}}$ & 0.9939$^{*\phantom{**}}$ & 0.9878$^{*\phantom{**}}$ & 0.9817$^{*\phantom{**}}$ & 0.9756$^{*\phantom{**}}$ \\

CBDC1 & 0.9934$^{***}$           & 0.9866$^{***}$           & 0.9793$^{**\phantom{*}}$ & 0.9672$^{\phantom{***}}$ & 0.9934$^{***}$           & 0.9868$^{***}$           & 0.9801$^{***}$           & 0.9735$^{***}$ \\

CBDC2 & 0.9938$^{***}$           & 0.9874$^{***}$           & 0.9802$^{***}$           & 0.9668$^{\phantom{***}}$ & 0.9940$^{***}$           & 0.9879$^{***}$           & 0.9819$^{***}$           & 0.9759$^{***}$ \\

CBDC3 & 0.9937$^{\phantom{***}}$ & 0.9872$^{\phantom{***}}$ & 0.9799$^{\phantom{***}}$ & 0.9672$^{\phantom{***}}$ & 0.9938$^{\phantom{***}}$ & 0.9876$^{\phantom{***}}$ & 0.9815$^{\phantom{***}}$ & 0.9753$^{\phantom{***}}$ \\

CBDC4 & 0.9937$^{\phantom{***}}$ & 0.9871$^{\phantom{***}}$ & 0.9797$^{\phantom{***}}$ & 0.9676$^{\phantom{***}}$ & 0.9938$^{\phantom{***}}$ & 0.9876$^{\phantom{***}}$ & 0.9814$^{\phantom{***}}$ & 0.9752$^{\phantom{***}}$ \\

\bottomrule
\multicolumn{9}{l}{\textit{Significance levels:} *** p<0.01, ** p<0.05, * p<0.1}
\end{tabular}

    \caption{Social welfare scores for the distribution of net wealth varying the conversion rule and different     inequality aversion parameter ($\epsilon$ and $\lambda$).}
    \label{NW}
    \end{table}

    \begin{table}[H]
    \scriptsize
    \centering
    \resizebox{\textwidth}{!}{\begin{tabular}{l|l|llll|llll}
\toprule
Scenario & & \multicolumn{4}{c|}{Atkinson} & \multicolumn{4}{c}{Mean-variance} \\
 & unemp & $\varepsilon=0.5$ & $\varepsilon=1.0$ & $\varepsilon=1.5$ & $\varepsilon=2.0$ & $\lambda=0.25$ & $\lambda=0.50$ & $\lambda=0.75$ & $\lambda=1.00$ \\
\midrule
Base & 0.0980 & 0.9937 & 0.9871 & 0.9797 & 0.9666 & 0.9938 & 0.9877 & 0.9815 & 0.9753 \\
$a_2=0.10$ & 0.0983 & 0.9938** & 0.9873** & 0.9801** & 0.9666 & 0.9939* & 0.9878* & 0.9817* & 0.9756* \\
$a_2=0.20$ & 0.0986*** & 0.9938** & 0.9873** & 0.9800** & 0.9673 & 0.9939* & 0.9878* & 0.9817* & 0.9756* \\
$a_2=0.30$ & 0.0987*** & 0.9938*** & 0.9874*** & 0.9802*** & 0.9668 & 0.9940*** & 0.9879*** & 0.9819*** & 0.9759*** \\
$a_2=0.40$ & 0.0993*** & 0.9939*** & 0.9875*** & 0.9804*** & 0.9685 & 0.9940*** & 0.9880*** & 0.9820*** & 0.9759*** \\
$a_2=0.50$ & 0.0996*** & 0.9939*** & 0.9874*** & 0.9804*** & 0.9666 & 0.9940*** & 0.9879*** & 0.9819*** & 0.9759*** \\
$a_2=0.60$ & 0.1007*** & 0.9938** & 0.9873** & 0.9802*** & 0.9684 & 0.9939 & 0.9877 & 0.9816 & 0.9755 \\
$a_2=0.70$ & 0.1017*** & 0.9937 & 0.9871 & 0.9800* & 0.9693*** & 0.9937** & 0.9874** & 0.9812** & 0.9749** \\
$a_2=0.80$ & 0.1041*** & 0.9934*** & 0.9866*** & 0.9793** & 0.9672 & 0.9934*** & 0.9868*** & 0.9801*** & 0.9735*** \\
\bottomrule
\multicolumn{10}{l}{\textit{Significance levels:} *** p<0.01, ** p<0.05, * p<0.1}
\end{tabular}
}
    \caption{Social welfare scores for the distribution of net wealth varying $a_2$ and different inequality aversion parameter ($\epsilon$ and $\lambda$).}
    \label{NW2}
    \end{table}

    The results suggest that a limited amount of CBDC (CBDC0, CBDC2-4 rules) does not harm the economy and may produce small welfare improvements. 
The welfare gains from the CBDC3 and CBDC4 rule are small and not statistically significant compared to the baseline model.
    When the degree of inequality aversion is high ($\epsilon=2$ or $\lambda=1$), the welfare improvement is limited and in most of the scenarios is not statistically significant. 
    Confirming the analysis in Section \ref{sec:results}, the smooth CBDC2 rule seems to perform the best in providing a social welfare improvement.

    Excluding values reported in the column for $\epsilon=2$, we always observe a lower social welfare for the CBDC1 rule compared to the no CBDC scenario. We can conclude that adoption of CBDC with a loose upper-bound harms the welfare of the economy.

    In Table \ref{NW2}, we investigate the optimal amount of CBDC varying $a_2$ in CBDC2. We allow a fixed amount of CBDC (10\% of deposits) and the possibility for consumers to substitute their deposits up to the fraction $a_2$ depending on the riskiness of the bank. 
    We report the social welfare scores varying $a_2 \in [0.1, \ 0.8]$ considering the distribution of net wealth. Notice that $a_4=0.8$.
    As $a_2$ increases, we observe that unemployment goes up. 
    According to the Atkinson and mean-variance social welfare functions, adoption of CBDC leads to a welfare improvement in most of the cases, unless the upper-bound rate to the conversion of CBDC  is above $0.6$.
    The upper-bound of the conversion rate that leads to the highest improvement of the social welfare function is around $0.4$. 

    \section{Acknowledgments}
    The support from the UKRI Grant entitled "The Large Agent Collider: Robust agent-based modelling as scale" awarded to Prof. Wooldridge (reference EP/W002949/1) is gratefully acknowledged. 
    In addition, the authors would like to acknowledge the use of the University of Oxford Advanced Research Computing (ARC) facility in carrying out this work. \url{http://dx.doi.org/10.5281/zenodo.22558}.  
    The authors acknowledge the support from the European Union - Next Generation EU - Project ‘GRINS - Growing Resilient, INclusive and Sustainable’ project (PE0000018); the National Recovery and Resilience Plan (NRRP) Spoke 4. The views and opinions expressed are only those of the authors and do not necessarily reflect those of the European Union or the European Commission. 
 
    \section{Conclusions}
    \label{sec:conclusion}
    Households may substitute deposits with CBDC that can be used for payments. The appealing feature for them is that, being a liability of the CB, CBDC is a safe asset. The chance to adopt it changes structurally financial intermediation in two directions: it reduces the deposits of banks and, therefore, their capability to lend to the economy, and it may exacerbate bank-runs in the case where risky banks ignite a flight-to-quality by households.
    Welfare and financial stability effects associated with CBDC have not been fully analysed mostly because the existing literature focuses on exogenous bank-runs. The agent-based model proposed in this paper permits us to consider together the decision to substitute deposits with CBDC and the riskiness of banks. 
   In this way, we are able to properly assess the implications of CBDC introduction for systemic financial stability and the potential for digital bank-runs.
    The paper contributes to the debate on CBDC in two ways: the assessment of disintermediation associated with the issuance of CBDC; and the optimal amount of CBDC.
    
    We show that the possibility of converting deposits into CBDC may ignite a flight-to-quality bank-run, if households are allowed to convert a large fraction of liquidity into CBDC. The effect on the economy is significant with lower growth and larger fluctuations. A 30\% deposit cap, alongside deposit insurance, would mitigate potential adverse effects on the broader economy.

    CBDC leads to a redistribution of wealth from firms to households, with a higher default rate of banks. Banks cope with agents' requests for liquidity by exchanging liquidity among themselves in the interbank market. 
    Macroeconomic and instability effects turn out to be significant only in the case where the maximum CBDC holding is set very high (80\% of deposits). 
    We find evidence that social welfare improves when the CBDC holdings are bounded at approximately 40\% of deposits. For larger holdings, the introduction of CBDC negatively affects the social welfare of the economy.

    We can conclude that the fear of disintermediation associated with CBDC is largely exaggerated. The effects on the real economy due to financial disintermediation can be effectively mitigated by imposing a reasonable cap to the adoption of CBDC. The economy turns out to be resilient to a non insignificant adoption of CBDC. 
    \vfill

    \pagebreak

    \footnotesize
    \bibliographystyle{apalike}
    \bibliography{biblio_cbdc}

\begin{thebibliography}{}

\bibitem[Adalid et~al., 2022]{ADAL}
Adalid, R., {\'A}lvarez-Bl{\'a}zquez, {\'A}., Assenmacher, K., Burlon, L., Dimou, M., L{\'o}pez-Quiles, C., Fuentes, N.~M., Meller, B., Mu{\~n}oz, M., Radulova, P., et~al. (2022).
\newblock Central bank digital currency and bank intermediation.
\newblock ECB Occasional Paper 2022/293, European Central Bank.

\bibitem[Agur et~al., 2022]{AGUR_ET_ALL}
Agur, I., Ari, A., and Dell’Ariccia, G. (2022).
\newblock Designing central bank digital currencies.
\newblock {\em Journal of Monetary Economics}, 125:62--79.

\bibitem[Ahnert et~al., 2023]{AHN}
Ahnert, T., Hoffman, P., Leonello, A., and Porcellacchia, D. (2023).
\newblock Central bank digital currency and financial stability.
\newblock Technical report, ECB Occasional Paper, 2023/2783.

\bibitem[Andolfatto, 2021]{AND}
Andolfatto, D. (2021).
\newblock Assessing the impact of central bank digital currency on private banks.
\newblock {\em The Economic Journal}, 131:525--540.

\bibitem[Assenmacher et~al., 2021]{ASSE}
Assenmacher, K., Berentsen, A., Brand, C., and Lamersdorf, N. (2021).
\newblock A unified framework for cbdc design: remuneration, collateral haircuts and quantity constraints.
\newblock Technical report, European Central Bank working paper series no.2578.

\bibitem[Atkinson et~al., 1970]{atkinson1970measurement}
Atkinson, A.~B. et~al. (1970).
\newblock On the measurement of inequality.
\newblock {\em Journal of economic theory}, 2(3):244--263.

\bibitem[Azzone and Barucci, 2023]{AZZ_BAR}
Azzone, M. and Barucci, E. (2023).
\newblock Evaluation of sight deposits and central bank digital currency.
\newblock {\em Journal of International Financial Markets, Institutions and Money}, 88:101841.

\bibitem[Barucci et~al., 2025]{BA_BR_MA}
Barucci, E., Brachetta, M., and Marazzina, D. (2025).
\newblock The adoption of central bank digital currency.
\newblock {\em Management Science}.

\bibitem[Beaumont, 2019]{beaumont2019approximate}
Beaumont, M.~A. (2019).
\newblock Approximate {B}ayesian computation.
\newblock {\em Annual review of statistics and its application}, 6(1):379--403.

\bibitem[Bindseil and Senner, 2024]{BIND2}
Bindseil, U. and Senner, R. (2024).
\newblock Destabilisation of bank deposits across destinations: assessment and policy implications.
\newblock Technical report, ECB working paper, 2024/2887.

\bibitem[Brei et~al., 2023]{brei2023effective}
Brei, M., Gambacorta, L., Lucchetta, M., and Parigi, B.~M. (2023).
\newblock How effective are bad bank resolutions? {N}ew evidence from {E}urope.
\newblock {\em Journal of Financial Stability}, 67:101153.

\bibitem[Brunnermeier and Niepelt, 2019]{BRUN_NIEP}
Brunnermeier, M.~K. and Niepelt, D. (2019).
\newblock On the equivalence of private and public money.
\newblock {\em Journal of Monetary Economics}, 106:27--41.

\bibitem[Burlon et~al., 2024]{BURL}
Burlon, L., Muñoz, M.~A., and Smets, F. (2024).
\newblock The optimal quantity of cbdc in a bank-based economy.
\newblock {\em American Economic Journal: Macroeconomics}, 16(4):172–217.

\bibitem[Cifuentes et~al., 2005]{SHIN}
Cifuentes, R., Ferrucci, G., and Shin, H.~S. (2005).
\newblock Liquidity risk and contagion.
\newblock {\em Journal of the European Economic Association}, 3(2/3):556--566.

\bibitem[Delli~Gatti et~al., 2011]{delli2011macroeconomics}
Delli~Gatti, D., Desiderio, S., Gaffeo, E., Cirillo, P., and Gallegati, M. (2011).
\newblock {\em Macroeconomics from the Bottom-up}.
\newblock Milano: Springer Milan.

\bibitem[Diamond and Dybvig, 1983]{DI_DY}
Diamond, D. and Dybvig, P. (1983).
\newblock Bank runs, deposit insurance, and liquidity.
\newblock {\em Journal of Political Economy}, 91:401--419.

\bibitem[Dyer et~al., 2024]{dyer2024black}
Dyer, J., Cannon, P., Farmer, J.~D., and Schmon, S.~M. (2024).
\newblock Black-box {B}ayesian inference for agent-based models.
\newblock {\em Journal of Economic Dynamics and Control}, 161:104827.

\bibitem[{European Banking Federation}, 2024]{EBF2024}
{European Banking Federation} (2024).
\newblock Banking in {Europe}: {EBF} facts \& figures 2024.
\newblock Technical report, European Banking Federation.
\newblock Accessed: 2025-02-10.

\bibitem[Fernández-Villaverde et~al., 2021]{FERN_ET_ALL}
Fernández-Villaverde, J., Sanches, D., Schilling, L., and Uhlig, H. (2021).
\newblock Central bank digital currency: Central banking for all?
\newblock {\em Review of Economic Dynamics}, 41:225--242.

\bibitem[Gertler and Kiyotaki, 2015]{GERT_KYI}
Gertler, M. and Kiyotaki, N. (2015).
\newblock Banking, liquidity, and bank runs in an infinite horizon economy.
\newblock {\em American Economic Review}, 105(7):2011–43.

\bibitem[Grazzini et~al., 2017]{grazzini2017bayesian}
Grazzini, J., Richiardi, M.~G., and Tsionas, M. (2017).
\newblock Bayesian estimation of agent-based models.
\newblock {\em Journal of Economic Dynamics and Control}, 77:26--47.

\bibitem[Gurgone and Iori, 2022]{gurgone2022macroprudential}
Gurgone, A. and Iori, G. (2022).
\newblock Macroprudential capital buffers in heterogeneous banking networks: insights from an abm with liquidity crises.
\newblock {\em The European Journal of Finance}, 28(13-15):1399--1445.

\bibitem[Gurgone et~al., 2018]{gurgone2018effects}
Gurgone, A., Iori, G., and Jafarey, S. (2018).
\newblock The effects of interbank networks on efficiency and stability in a macroeconomic agent-based model.
\newblock {\em Journal of Economic Dynamics and Control}, 91:257--288.

\bibitem[Herbst and Schorfheide, 2016]{herbst2016bayesian}
Herbst, E.~P. and Schorfheide, F. (2016).
\newblock {\em Bayesian estimation of DSGE models}.
\newblock Princeton University Press.

\bibitem[Keister and Monnet, 2022]{KEI_MON}
Keister, T. and Monnet, C. (2022).
\newblock Central bank digital currency: Stability and information.
\newblock {\em Journal of Economic Dynamics and Control}, 142:104501.

\bibitem[Keister and Sanches, 2023]{KEI_SAN}
Keister, T. and Sanches, D. (2023).
\newblock Should central banks issue digital currency?
\newblock {\em The Review of Economic Studies}, 90(1):404--431.

\bibitem[Kim and Kwon, 2023]{KIM_KWON}
Kim, Y.~S. and Kwon, O. (2023).
\newblock Central bank digital currency, credit supply, and financial stability.
\newblock {\em Journal of Money, Credit and Banking}, 55(1):297--321.

\bibitem[Lenormand et~al., 2013]{lenormand2013adaptive}
Lenormand, M., Jabot, F., and Deffuant, G. (2013).
\newblock Adaptive approximate {B}ayesian computation for complex models.
\newblock {\em Computational Statistics}, 28(6):2777--2796.

\bibitem[Lux, 2022]{lux2022bayesian}
Lux, T. (2022).
\newblock Bayesian estimation of agent-based models via adaptive particle {Markov} chain {Monte} {Carlo}.
\newblock {\em Computational Economics}, 60(2):451--477.

\bibitem[Nyffenegger, 2024]{NYFFE}
Nyffenegger, R. (2024).
\newblock Central bank digital currency and bank intermediation: Medium of exchange vs. savings vehicle.
\newblock {\em European Economic Review}, 170:104890.

\bibitem[Platt, 2020]{platt2020comparison}
Platt, D. (2020).
\newblock A comparison of economic agent-based model calibration methods.
\newblock {\em Journal of Economic Dynamics and Control}, 113:103859.

\bibitem[Popoyan et~al., 2017]{lilit2016taming}
Popoyan, L., Napoletano, M., and Roventini, A. (2017).
\newblock Taming macroeconomic instability: Monetary and macro-prudential policy interactions in an agent-based model.
\newblock {\em Journal of Economic Behavior \& Organization}, 134:117--140.

\bibitem[Varaart, 2025]{veraart2025}
Varaart, L. A.~M. (2025).
\newblock Modelling contagious bank runs.

\bibitem[Williamson, 2022a]{WILL2}
Williamson, S. (2022a).
\newblock Central bank digital currency: Welfare and policy implications.
\newblock {\em Journal of Political Economy}, 130(11):2829--2861.

\bibitem[Williamson, 2022b]{WILL}
Williamson, S.~D. (2022b).
\newblock Central bank digital currency and flight to safety.
\newblock {\em Journal of Economic Dynamics and Control}, 142:104146.

\bibitem[Zachary et~al., 2025]{feinstein2025}
Zachary, F., Grzegorz, H., and Andreas, S. (2025).
\newblock The not-so-hidden risks of 'hidden-to-maturity' accounting: on depositor runs and bank resilience.

\end{thebibliography}
    \clearpage

    \begin{subappendices}
    \section{Appendix}\label{sec.appendix}
 
    \normalsize
    \subsection{Model calibration}\label{sec.calibration}
    \subsubsection{Number of agents}\label{sec.number_agents}
    The number of agents is calibrated on the statistics for the 27 countries belonging to the European Union (EU). 
    Eurostat reports 32,721,956 enterprises operating in the EU in 2023, with a ratio of 5 persons employed per enterprise.\footnote{Enterprise statistics by size class and NACE Rev. 2 activity (from 2021 onward), \url{https://doi.org/10.2908/SBS_SC_OVW}} The number of employees per worker was approximately 0.84 in 2022. 
    For the sake of simplicity, we assume all employed workers are employees and that one household in the model corresponds to one employee. Guided by these ratios, we set the number of agents to 500 firms and 2500 households. The cardinality aims to reduce the computational burden while maintaining a reasonable number of agents to ensure meaningful interactions in the markets. Data from the \citet{EBF2024} show that, at the end of 2023, there were 4,297 credit institutions, which means that the ratio of enterprises per credit institution was around 7,615. According to these ratios, considering 10 banks, we should scale up the number of firms to approximately 76,000 and the number of households to about 380,000. This would transform the model to a large-scale one, which is outside the scope of this work and would require a substantial increase in computational power and modifications to the code. We assume a 50:1 ratio, resulting in a number of 10 banks which is enough to ensure a meaningful interaction among them, especially in the interbank market, and among banks, firms, and households. 
    
    \subsubsection{Initial values}\label{sec.initial_values}
    The initial net wealth of the agents is derived from the data of the 20 countries in the Euro area.\footnote{We consider 20 countries instead of 27 in the EU as aggregate monetary data are only available for the 20 country in the currency union in 2024.} The net wealth of households and firms, which corresponds to deposits by assuming no liabilities, is determined from the ratios to the nominal GDP of deposits of employees and deposits of non-financial corporations. We first compute the actual ratios of deposits to GDP based on the data from the European Central Bank (ECB), which are, respectively, around 1.06 for employees and 0.90 for non-financial corporations in 2024.\footnote{
    See \url{data.ecb.europa.eu}
    \begin{itemize}
        \item Series key: \url{DWA.Q.I9.S14.A.LE.F2M.WSE.EUR.S.N} (deposits of employees)
        \item Series key: \url{QSA.Q.N.I9.W0.S11.S1.N.A.LE.F2M.T._Z.XDC._T.S.V.N._T} (deposits placed by non-financial corporations)
        \item Series key: \url{MNA.Q.Y.I9.W2.S1.S1.B.B1GQ._Z._Z._Z.EUR.V.N} (GDP at market prices).
    \end{itemize}
    }
    Then, we set the initial values of deposits by multiplying these ratios by the initial value of potential GDP in the model.\footnote{
    Potential GDP is given by the product of the initial values of the price mark-up $1+\mu_0$, the wage rate $W_0$, the number of households $N^H$, one minus a measure of the natural unemployment rate $u^{\star}$, and the maximum quantity of labour supplied by each household normalized to 1.
    $$ GDP_0 \equiv P_0Y_0 = (1+\mu_0) W_0 (1-u^{\star}) N^H.$$
    } 
    The net wealth of banks is set at $10\%$ of the sum of households' and firms' deposits, i.e., the total deposits held by banks, drawn from the supervisory banking statistics for significant credit institutions directly supervised by the ECB.\footnote{
    See statistics on balance sheet composition and profitability of banks in the Euro area \url{https://data.ecb.europa.eu/main-figures/supervisory-banking-data/balance-sheet-composition-and-profitability}.
    }
    Following this approach, the wage rate at time 0 ($W_0$) works as a numéraire upon which the other monetary quantities scale. 
    By design, the total net wealth of agents corresponds to the total quantity of money in the system. Therefore, $W_0$ determines the nominal money balance and the aggregate stock of money in the economy.
    
    \subsubsection{Labour market}\label{par.calibration_labour_mkt}
    The unemployment rate $u^{\star}$ is the long-term average calibrated on data for the Euro area, see Table \ref{tab.estimation_targets}, and it is assumed to correspond to the natural rate of unemployment.
    
    The mechanism for transitioning out of involuntary unemployment described in Section \ref{sec.labour} is modelled through a binomial model in which job seekers have a probability $p(s)$ of a successful match with a firm in $n$ trials: 
    \[
        p(s)=\frac{n!}{k!(n-k)!}p^k(1-p)^{n-k}\;\;. 
    \]

    We set $n=2$, $k=1$, and $p=0.5$, resulting in $p(s)=0.5$ to match the quarterly transition probability from unemployment observed in  the EU labour markets, assuming that unemployed can only move to employment but cannot leave the labour force.\footnote{See labour market flow statistics, \url{      https://ec.europa.eu/eurostat/statistics-explained/index.php?    title=Labour_market_flow_statistics_in_the_EU}.}

    \subsubsection{Calibrated model parameters}\label{sec.calibrated_parameters}
    Parameter values are reported in Table~\ref{tab.parameters}. 
    	\begin{table}[h!]
		\centering
		\scriptsize
		\begin{tabular}{lrl}
			\toprule
			Parameter & Value & Description\\ 
			\midrule
   			$T$  & $1,000$ & Time length of the simulation \\
                $N^H$ & $2,500$ & Number of households (section \ref{sec.number_agents}) \\
                $N^F$ & $500$ & Number of firms (section \ref{sec.number_agents})\\
                $N^B$ & $10$ & Number of banks (section \ref{sec.number_agents})\\
                $\theta^H$ & $0.3$ & Household income tax rate \eqref{eq.nw_H}\\
                $\theta^F$ & $0.3$ & Tax rate for firms \eqref{eq.nw_F_dynamics}\\
                $\theta^B$ & $0.3$ & Tax rate for banks \eqref{eq.deltanwB} \\
                $d^F$ & $0.06^\dag$ & Variable component in firms' dividends \eqref{eq.nw_F_dynamics}\\
                $\delta^F$ & $0.25^\dag$ & Dividend share of firms \eqref{eq.nw_F_dynamics}\\
                $\delta^B$ & $0.49^\dag$ & Dividend share of banks \eqref{eq.deltanwB}\\
                $\alpha$ &  $1$ & Labour productivity \eqref{eq.labour_target} \\ 
			$c_1$  & $0.8$ & Marginal propensity to consume out of income \eqref{eq.C} \\ 
			$c_2$ & $0.2$ & Marginal propensity to consume out of wealth \eqref{eq.C} \\ 
                $n^{cred}$ & 3 & Maximum number of attempts to borrow on the credit market  (section \ref{sec.matching})\\
                $n^{ibtent}$ & 5 & Maximum number of attempts to borrow on the interbank market (section \ref{sec.matching})\\
                $n^{tent}$ & 2 & Number of times consumers visit the goods market (section \ref{sec.matching})\\
                $r^L$ & 0.03 & Lower bound of interest rate, paid by the Central Bank on bank reserves (yearly) (section \ref{sec:banks}) \\
                $r^D$ & 0.03 & Interest rate paid by banks on deposits (yearly) (section \ref{sec:banks}) \\
                $r^B$ & 0.03 & Interest rate paid by government on bonds (yearly) (section \ref{sec.GOV}) \\
                $r^{CBDC}$ & 0.03 & Interest rate paid by central bank on CBDC (yearly) (section \ref{sec.GOV}). \\
                $r^H$ & 0.04 & Upper bound of the interest rate (yearly) (section \ref{sec:banks}) \\
                $rr$ & 0.10 & Regulatory reserve ratio, (section \ref{sec.banks_balance_sheet}) \\
                $v_0$ & $1-\frac{1+r^L}{1+r^H}$ & Composite parameter \eqref{eq.rho_f}, \eqref{eq:def_prob_ib}\\
                $v_1$ & $2^\dag$ & Sensitivity of $r^f$ to default probability \eqref{eq.rho_f}, \eqref{eq:def_prob_ib}\\
                $l^\star$ & $4.4^\dag$ & Scale parameter for leverage \eqref{eq.rho_f} \\
                $l^{b\star}$ & 2 & Scale parameter for leverage \eqref{eq:def_prob_ib}\\
                $\lambda$ & 1/0.07 & Inverse CET1 ratio (regulatory leverage ratio) \eqref{eq:loan_supply} \\
                $\omega_1$ & 1 & Risk weight on loans to firms \eqref{eq:loan_supply} \\
                $\omega_2$ & 0.3 & Risk weight on interbank loans \eqref{eq:loan_supply} \\  
                $matur$ & 1 & Loan maturity \\
                $\tau$ & 20 & Length of firms' and banks' memory and banks' VaR horizon\\
                $tail$ & 0.99 & VaR tail probability (section \ref{sec.credit_supply}) \\
                $p_0$ & 1 &  Asset price (bonds and loans) at the beginning of each time unit \eqref{sec.liquidation}\\
                $\epsilon^{loans}$ & -0.9 & Price elasticity of loans\\
                $\epsilon^{bonds}$ & -1.5 & Price elasticity of bonds\\
                $a^L$ & 0.8 & Weight of current vs past loans in $L^E$ (section \ref{sec.ib_market})\\
                $\gamma_q$ & 0.1 & Threshold for quantity adjustment \eqref{eq.Y_adustment}, \eqref{eq.mark-up} \\
                $\gamma_p$ & $0.87^\dag$ & Threshold for price adjustment 
                \eqref{eq.Y_adustment}, \eqref{eq.mark-up} \\
                $q_b$ &  $0.4^\dag$ & Quantity adjustment, upper bound \eqref{eq.Y_adustment} \\
                $\mu_{max}$ & 0.25 & Maximum mark-up \eqref{eq.mark-up} \\ 
                $\mu_{min}$ & 0.01 & Minimum mark-up  \eqref{eq.mark-up} \\ 
                $\mu_a$ & $0.78^\dag$ &  Mark-up adjustment, upper bound \eqref{eq.mark-up} \\
                $\mu_0$ & $0.19^\dag$ & Initial mark-up on prices \eqref{eq.price} \\
                $w_b$ & $0.01^\dag$ & Wage rate adjustment, upper bound \eqref{eq.Phillips}\\
                $u^{\star}$ & $0.094$ & Long-term rate of unemployment \eqref{eq.Phillips}\\
                $F_h$ & $0.3^\dag$ & Share of firms observed by a consumer on the goods market (section \ref{sec.matching})\\                
                $\text{bid}_b$ & $0.15$ & Mark-up adjustment on the interbank bid rate, upper bound \eqref{eq:bid_rate} \\
                $recap^F$ & $2$ & Minimum time between default and recapitalization (firms) (section \ref{sec.defaults})\\
                $recap^B$ & $4$ & Minimum time between default and recapitalization (banks) (section \ref{sec.defaults}) \\
                $\phi_{ext}$ & $0.1$ & Share of banks' deposits invested in bonds (section \ref{sec.banks_balance_sheet})\\
                $\varsigma$ & $0.15$ & Maximum equity loss for a lender \eqref{eq:loan_exposure}\\
                $\zeta$ & $0.16^\dag$ & Priority given to internal finance over borrowing \eqref{eq.loantarget} \\
                $RM^{*}$ & $6$ & Threshold risk measure for CBDC \eqref{eq:cbdc2} \\
                $RM^{lim}$ & $7.6$ & Maximum leverage for CBDC \eqref{eq:cbdc3} \\
                $a1$ & 0.1 & CBDC share of deposits, \eqref{eq:cbdc1}\\
                $a2$ & 0.3 & CBDC share of deposits, \eqref{eq:cbdc2} \\
                $a3$ & 0.7 & CBDC share of deposits, \eqref{eq:cbdc3} \\
                $IT^{*}$ & $5.4$ & Insured value of deposits \eqref{eq:cbdc4} \\                
 			\bottomrule 
            \end{tabular}
            \vspace{1ex}
            \begin{minipage}{\textwidth}
            \centering
            \end{minipage}
            \caption{Model parameters and calibrated values. Those values estimated by Bayesian inference are denoted by $\dag$.}
            \label{tab.parameters}	
	\end{table}

    \FloatBarrier

    \subsection{Bayesian estimation}\label{sec.estimation}
    The intrinsic complexity of agent-based models makes it challenging to derive and estimate closed-form solutions of the model and requires simulation-based techniques to match real-world data. \citet{platt2020comparison} shows that Bayesian estimation can outperform frequentist approaches for the calibration of these models. Bayesian methods are also commonly used in macroeconomics for the estimation of Dynamic Stochastic General Equilibrium (DSGE) models, see \citep{herbst2016bayesian}, and are increasingly adopted in the estimation of agent-based models, see \citep{grazzini2017bayesian, lux2022bayesian, dyer2024black}. 
    
    In our calibration, we use Bayesian inference to fit the model to the data. In other words, while some parameters in Table \ref{tab.parameters} are set from the data, such as the number of households and the interest rates, other parameters, e.g. those governing the adjustment of price and production, are estimated by Bayesian inference in order to match the summary statistics of selected real-world time series.
   
    We focus on capturing the first two statistical moments of selected time series through Bayesian inference to find a reasonable match between key simulated time series and their real-world counterparts, thus restricting the estimation scope. The description of real-world time series with their target moments is presented in Table \ref{tab.estimation_targets}. Means and standard deviations are computed over the same time range in which bank default data are available, except for the CET1 ratio, which is unavailable before 2015. 

    We cannot use Bayesian estimation to match the default rate of banks in Table \ref{tab.estimation_targets} as the standard deviation is not available from \citet{brei2023effective}. Still, reproducing plausible summary statistics for the the variable is relevant to this study. Therefore, we match ex post its first moment by choosing an appropriate combination of the estimated parameters among those in the acceptable set that produces a mean close to the simulated series (see Table \ref{tab:RMSD}). 

        \begin{table}[ht!]
    \centering
    \footnotesize
    \begin{tabular}{lrrrrr}
    \toprule
     Series & Mean & Std & Range & Source\\ 
    \midrule
     Unemployment rate & 0.095 & 0.013 &  2000Q1-2019Q4 & \url{data.ecb.europa.eu}\tablefootnote{Series key: \url{LFSI.Q.I9.S.UNEHRT.TOTAL0.15_74.T}}\\ 
     Inflation rate (demeaned)  & 0 & 0.009 & 2000Q1-2019Q4 & \url{data.ecb.europa.eu}\tablefootnote{Series key: \url{ICP.M.U2.N.000000.4.INX}}\\
     Credit to GDP & 0.713 & 0.064 & 2000Q1-2019Q4 & \url{data.ecb.europa.eu}\tablefootnote{Series key: \url{QSA.Q.N.I9.W0.S11.S1.C.L.LE.F3T4.T._Z.XDC_R_B1GQ_CY._T.S.V.N._T}} \\ 
     CET1 ratio & 0.150 & 0.007 & 2015Q2-2019Q4 & \url{data.ecb.europa.eu}\tablefootnote{Series key: \url{SUP.Q.B01.W0._Z.I4002._T.SII._Z._Z._Z.PCT.C}}\\
     Default rate of banks & 0.009 & na & 2000-2019 & \citet{brei2023effective}, Table 1\tablefootnote{We consider the ratio of recapitalized and restructured banks to the total number of banks. The standard deviation of real-world data is not available (na).}\\
    \bottomrule 
    \end{tabular}
    \caption{Target moments for the Bayesian estimation. Mean and standard deviations refer to quarterly data. Inflation is computed year-over-year. Standard deviation for the default rate of banks is not available (na).} \label{tab.estimation_targets}
    \end{table}
    \FloatBarrier

    Bayesian inference aims to find the posterior distribution of the parameters. By Bayes' theorem, the posterior is proportional to the likelihood multiplied by the prior
    $$ p(\boldsymbol{\theta}|\mathbf{y}) = \frac{p(\mathbf{y}|\boldsymbol{\theta})p(\boldsymbol{\theta})}{p(\mathbf{y})} \propto p(\mathbf{y}|\boldsymbol{\theta}) p(\boldsymbol{\theta})\;\;,$$
    where $\boldsymbol{\theta}$ is an unknown parameter vector; 
    $p(\boldsymbol{\theta}|\mathbf{y})$ is the posterior distribution of $\boldsymbol{\theta}$ given the observed data 
    $\mathbf{y}= {y_1, \cdots, y_n}$; $p(\boldsymbol{\theta})$ is the prior distribution of the parameters $\boldsymbol{\theta}$, namely the belief about the distribution before observing the data; 
    $p(\mathbf{y}) = \int p(\mathbf{y} | \boldsymbol{\theta}) p(\boldsymbol{\theta}) d\boldsymbol{\theta}$ is the the marginal probability of $\mathbf{y}$. The likelihood function $p(\mathbf{y}|\boldsymbol{\theta})$ represents the conditional probability of $\mathbf{y}$ given that the parameter vector $\boldsymbol{\theta}$ is true.

    Although conceptually simple, estimation of the likelihood presents significant computational obstacles. Instead of computing or estimating the likelihood function, we use an estimation algorithm that produces an approximate estimate of the posterior distribution of parameters. In other words, we resort to a likelihood-free method belonging to the Approximate Bayesian Computation (ABC) family \citep[for an overview on ABC, see][]{beaumont2019approximate}.
     
    Instead of evaluating likelihood, ABC methods use the concept of rejection sampling. A set of parameters drawn from the prior is accepted if the distance between summary statistics of some kind of simulated and real data falls below a given threshold and is rejected otherwise. In other words, if the model outcome is close enough to the data, the set of parameters is deemed satisfactory to approximate the true posterior and so to reproduce the target data via ABM simulations.

    The chosen summary statistic is the mean of simulated time series, computed after removing the first 500 transient periods. The distance function $d$ measures the deviation between the summary statistics of the simulated time series and the real data. For this estimation problem, we use the Euclidean norm of the difference between the mean of the simulated time series ($\mu_{sim}$) and the mean of real data ($\mu_{obs}$), normalized by the standard deviation of real data ($\sigma_{obs}$):
    $$ d =  \left[\sum_k^K \left(\frac{\mu_{sim,k} - \mu_{obs,k}}{\sigma_{obs,k}}\right)^{2} \right]^{1/2}\;\;.$$

    Computing the posterior distribution requires sampling many times from the prior and retaining those samples whose distance falls below a tolerance threshold. The strategy for computing the posterior is known as the sampling scheme. We implement an efficient sampling scheme based on the algorithm proposed in \citet{lenormand2013adaptive}, defined as Adaptive Population Monte Carlo (APMC). In a nutshell, APMC refines the classical Sequential Monte Carlo (SMC) scheme by producing unbiased estimates of the posterior, providing an adaptive tolerance threshold and a stop criterion, and reducing the computational cost compared to other SMC schemes.\footnote{The authors show that the performance of APMC is higher when compared to a Population Sequential Monte Carlo, Replenishment Sequential Monte Carlo, Adaptive Sequential Monte Carlo, and a Rejection-based ABC.}

    In our implementation, the key steps of the algorithm are as follows.
    \begin{enumerate}[itemsep=0.05cm]
        \item[0.] Set the initial values: number of combinations of parameter values  (particles) $N=200$, 
        weights of all particles $\omega_i = 1/N,\; 1\leq i\leq N $, minimum acceptance level $\rho_{acc_{min}}=0.05$, and the quantile $\alpha=0.5$.
        \item Sample $N$ particles  $(\theta)_{i=1,\dots,N}$ from the prior distribution $p(\mathbf{\theta})$ with a Latin hypercube.
        \item For each particle, simulate $MC$ replicates of the model (Monte Carlo) of length $T=1,000$ under a set of random seeds of length $MC$.
        \item Discard burn-in time $T_{burnin}=500$ from the simulated  series, merge them across Monte Carlo simulations, and compute summary statistics.
        \item Compute the distance $d$ between summary statistics from the simulated and real data.
        \item Retain the $\alpha N$ particles whose distance is below a tolerance threshold defined as the $\alpha$-quantile of distances.
        \item Resample $N - \alpha N$ new particles from a Gaussian probability density function, whose parameters are estimated from the retained particles and assign them weights.
        \item Concatenate the retained particles $\alpha N$ and the newly sampled particles $N - \alpha N$
        \item Repeat 2.-7.
        \item Stop when the proportion of newly sampled particles $N - \alpha N$ whose distance is below the threshold level $\varepsilon$ is less than the minimum acceptance level $\rho_{acc_{min}}$ or the maximum number of iterations is reached.
    \end{enumerate}

\begin{figure}[h]
    \centering
    \includegraphics[width=1\linewidth]{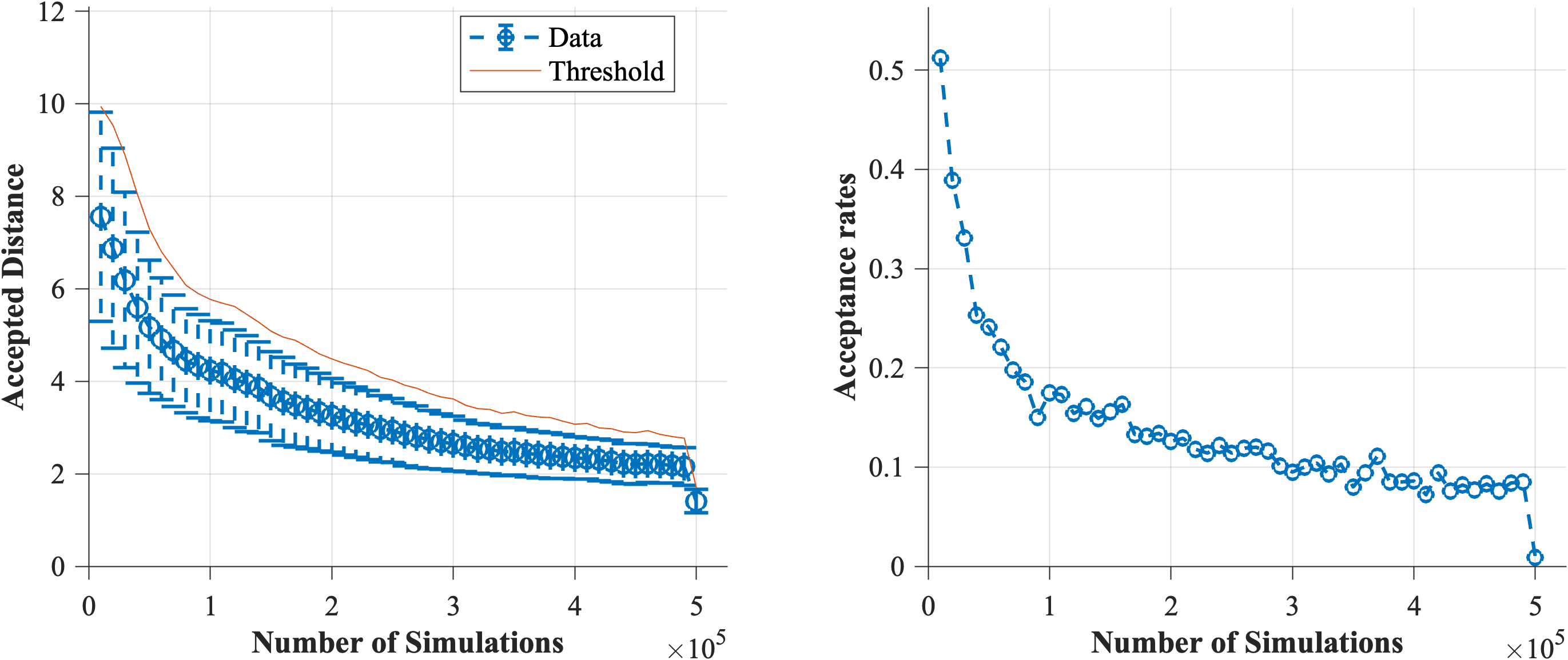}
    \caption{Accepted distances (left) and acceptance rates (right) for the Adaptive Population Monte Carlo estimation.}
    \label{fig:posterior_qual}
    \end{figure}

    We start from a set of $N=2,000$ particles for 14 free parameters. Each particle is simulated $MC=5$ times. The APMC algorithm is repeated 50 times, resulting in a total number of simulations of 500,000. The computations were performed using MATLAB's parallel \textit{parfor} loop on a 48 multicore high-performance computing cluster, for which we acknowledge the use of the University of Oxford Advanced Research Computing (ARC) facility in carrying out this work. The accepted distances and the minimum acceptance rates in simulations are represented in Figure \ref{fig:posterior_qual}. 

    \FloatBarrier
    \FloatBarrier

\begin{figure}[h]
    \centering
    \includegraphics[width=1\linewidth]{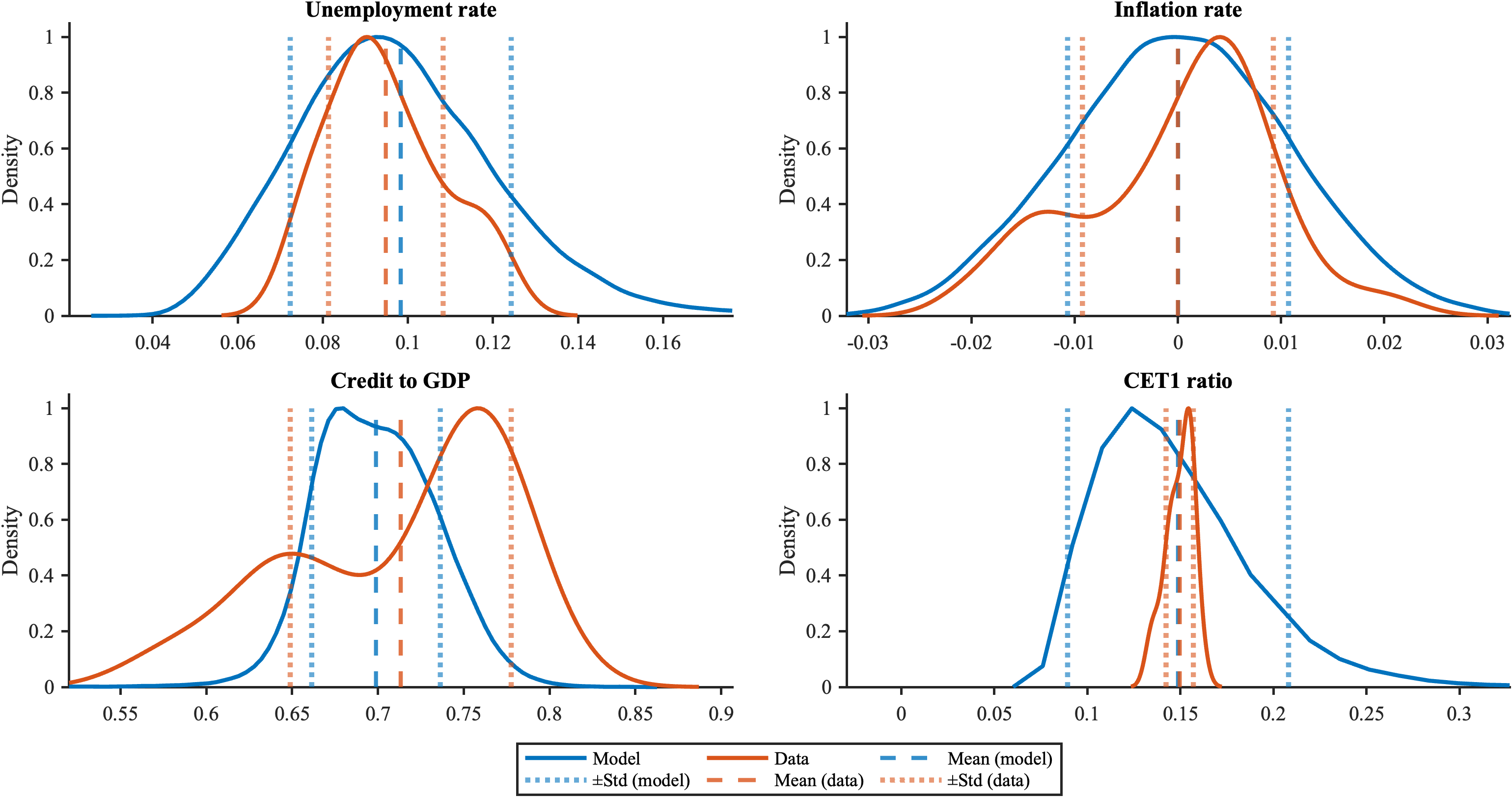}
    \caption{
    Kernel density estimates for probability distribution functions simulated series (Model, blue) and real-world data (Data, red). Means (dashed lines) and standard deviations (dotted lines) are shown in their respective colours.}
    \label{fig:estimation_densities}
    \end{figure}

The estimation results satisfactorily replicate the key properties observed in real-world data.
    The root mean square deviations (RMSD) for the mean and standard deviation from the target real-world series are reported in Table \ref{tab:RMSD}, where the simulated model uses the estimated parameters. 
    Kernel density estimates of the probability distribution function of simulated and real-world series are in Figure \ref{fig:estimation_densities}. 
    The simulated time series show small deviations from the target series for mean and the standard deviation of the unemployment and inflation rate, while the standard deviation of the simulated CET1 ratio is around one order of magnitude above the target.     
    The estimated parameters are reported in Table \ref{tab.parameters}.

    \FloatBarrier
    \begin{table}[h]
\centering
\footnotesize
\begin{tabular}{l|rrr|rrr}
\toprule
 & \multicolumn{3}{c|}{\textbf{Mean}} & \multicolumn{3}{c}{\textbf{Std}} \\
\textbf{Variable} & Simulated & Target & RMSD & Simulated & Target & RMSD \\
\midrule
Unemployment rate & 0.098 & 0.095 & 0.003 & 0.026 & 0.013 & 0.013 \\
Inflation rate & 0.000 & 0.000 & 0.000 & 0.011 & 0.009 & 0.001 \\
Credit to GDP & 0.699 & 0.713 & 0.014 & 0.037 & 0.064 & 0.027 \\
CET1 ratio & 0.149 & 0.150 & 0.001 & 0.059 & 0.007 & 0.052 \\
Bank default rate & 0.012 & 0.007 & 0.005 & 0.048 & 0.004 & 0.044 \\
\bottomrule
\end{tabular}
\caption{Root Mean Square Deviation (RMSD) for mean and standard deviation of simulated time series compared to real-world targets. Statistics are computed on 100 replicates.}
\label{tab:RMSD}
\end{table}

    \FloatBarrier
    \FloatBarrier

    \subsection{Accounting}\label{sec:accounting}
 The model’s equations are classified into behavioural and accounting equations.
    The model includes both \emph{stock} and \emph{flow} variables. Consistency across the equations is verified through a stock-flow consistent accounting system. It is composed of a \emph{transactions flow matrix} (Tab.~\ref{tab:balancesheet})  and a \emph{balance sheet matrix} (Tab.~\ref{tab:transactions}). The former captures the changes in stock variables between the beginning and the end of each period, while the latter shows the level of these variables at a given point in time, providing an accountancy-based representation of the model.
    
    \subsubsection{Aggregate balance sheet and transactions matrix}\label{sub.agg_bal and trans_mat}
    Table \ref{tab:balancesheet} represents the aggregate balance sheet of the economic system. Each row and column sums to zero, ensuring that every transaction recorded for one class of agents is offset by an equivalent entry for the corresponding counterpart.

    Since it is assumed that (i) there is no physical capital and (ii) inventories are perishable, the firms' accounts sum to zero and the sum of all the net wealth is zero, so that the government has a negative net wealth:\footnote{In a model with physical capital and/or inventories the sum of the latter plus the sum of the net wealth should be zero.}
    
    \[\sum_{i\in N^H}nw^{H}_i+\sum_{j\in N^F}nw^{F}_j+\sum_{h\in N^B}nw^{B}_h+nw^{G}=0\;\;.\]
	
    Table \ref{tab:transactions} captures the aggregate exchanges taking place in the system. Each flow originates from one class of agents and is directed toward another, while intra-class flows are omitted at the aggregate level. These flows are reported in the rows of the matrix.
    The aggregate flows associated with each class of agents are represented in the columns and can be divided into current accounts (CA) and capital accounts (KA). The current account records the inflows and outflows arising from payments and earnings, whereas the capital account captures changes in the agents’ balance sheets, that is, variations in their assets and liabilities.
    \bigskip
    \FloatBarrier
    
    \begin{minipage}{\linewidth}
    \scriptsize{}
    \centering
    \begin{tabularx}{\textwidth}{>{\raggedright\arraybackslash}X 
                                >{\centering\arraybackslash}X 
                                >{\centering\arraybackslash}X 
                                >{\centering\arraybackslash}X 
                                >{\centering\arraybackslash}X 
                                >{\centering\arraybackslash}X 
                                >{\centering\arraybackslash}X}
        \toprule
        & HH  & FF & BB  & CB & Gov & $\sum$ \\ 
        \midrule
        \addlinespace
        Deposits   & $+D^{H}$  & $+D^{F}$ & $-D^B$ &  &  & 0 \\ 
        Loans      &           & $-L^F$   & $+L^F$  &  &  & 0 \\ 
        Bonds      &           &          &  $+B^B$   & $+B^{CB}$ &$-B$ & 0 \\ 
        Reserves   &           &          & $+R$  &  $-R$ &  & 0 \\ 
        CBDC       & $+CBDC$   &          &        & $-CBDC$ &  & 0 \\
        \midrule
        Net Worth  & $-nw^{H}$ & $-nw^{F}$ & $-nw^{B}$  & $-nw^{CB}$ & $-nw^{G}$ & 0 \\ 
        \midrule 
        $\sum$     & 0 & 0 & 0 & 0 & 0 & 0 \\ 
        \bottomrule
    \end{tabularx} 
        \captionof{table}{\small{Aggregate Balance Sheet}} \label{tab:balancesheet}
    \vspace{0.2cm} 
    \scriptsize{\textit{Variables measured at current prices. Assets (+), liabilities (-).}\\
    HH = Households, FF = Firms, BB = Banks, CB = Central Bank, Gov = Government.}
\end{minipage}

    \begin{minipage}{\linewidth}
    \scriptsize{}
    \centering
    \begin{tabularx}{\textwidth}{>{\raggedright\arraybackslash}p{3.2cm} 
                                >{\centering\arraybackslash}X 
                                >{\centering\arraybackslash}X 
                                >{\centering\arraybackslash}X 
                                >{\centering\arraybackslash}X 
                                >{\centering\arraybackslash}X 
                                >{\centering\arraybackslash}X 
                                >{\centering\arraybackslash}X 
                                >{\centering\arraybackslash}X 
                                >{\centering\arraybackslash}r}
        \toprule
        & HH  & \multicolumn{2}{c}{FF} & \multicolumn{2}{c}{BB} & \multicolumn{2}{c}{CB} & Gov & $\sum$ \\ 
        \cmidrule(lr){3-4} \cmidrule(lr){5-6} \cmidrule(lr){7-8}
        &  & CA & KA & CA & KA & CA & KA &  & \\ 
        \midrule
        Consumption      & $-C$  & $+C$  &  &  &  &  &  &  & 0 \\ 
        Transfers       & $+G$  &       &  &  &  &  &  & $-G$ & 0 \\ 
        Production      &       & $+PY$   &  &  &  &  &  &  & 0 \\ 
        Wages          & $+WN$ & $-WN$ &  &  &  &  &  &  & 0 \\ 
        Taxes          & $-T^H$ & $-T^F$ &  & $-T^B$ &  &  &  & $+T$ & 0 \\ 
        Profits Firms  & $+\delta\Pi^{F}$ & $-\Pi^{F}$ & $+(1-\delta)\Pi^{F}$ &  &  &  &  &  & 0 \\ 
        Profits Banks  & $+\delta\Pi^{B}$ &  &  & $-\Pi^{B}$ & $+(1-\delta)\Pi^{B}$ &  &  &  & 0 \\ 
        Profits CB     &  &  &  &  &  & $-\Pi^{CB}$ &  & $+\Pi^{CB}$ & 0 \\ 
        \addlinespace
        Deposits Interest  & $+r^{D}D^{H}$ & $+r^{D}D^{F}$ &  & $-r^{D}D$ &  &  &  &  & 0 \\ 
        Loans Interest     &  & $-r^{f}{L}^f$ &  & $+r^{f}{L}^f$ &  &  &  &  & 0 \\
        Bonds Interest     &  &  &  &  &  & $+r^{B}B$ &  & $-r^{B}B$ & 0 \\ 
        Reserves Interest  &  &  &  & $+r^{R}R$ &  & $-r^{R}R$ &  &  & 0 \\ 
        \addlinespace
        $\Delta$ Loans      &  &  & $+\Delta L$ &  & $-\Delta L$ &  &  &  & 0 \\ 
        $\Delta$ Bonds      &  &  &  &  &  &  & $-\Delta B$ & $+\Delta B$ & 0 \\ 
        $\Delta$ Reserves   &  &  &  &  & $-\Delta R$ &  & $+\Delta R$ &  & 0 \\ 
        $\Delta$ Deposits   & $-\Delta D^{H}$ &  & $-\Delta D^{F}$ &  & $+\Delta D$ &  &  &  & 0 \\ 
        \midrule
        $\sum$             & 0 & 0 & 0 & 0 & 0 & 0 & 0 & 0 & 0 \\ 
        \bottomrule
    \end{tabularx}
        \captionof{table}{\small{Aggregate Transactions Flow Matrix}} \label{tab:transactions}

    \vspace{0.2cm} 
    \scriptsize{\textit{Variables measured at current prices. Sources of funds (+), uses of funds (-).}\\
    HH = Households, FF = Firms, BB = Banks, CB = Central Bank, Gov = Government.}
\end{minipage}

    \FloatBarrier
    \clearpage

    \subsection{Matching and networks structures}\label{sec.matching_and_networks}
    \subsubsection{The matching mechanism}\label{sec.matching}

    A matching mechanism governs interactions in the goods market. At each time step, households observe a randomly selected subset of firms, rank them in ascending order based on prices, and allocate their consumption budget prioritising the lowest-priced firms. This process, which can be repeated $n^{tent}$ times by each household, continues until either the consumption budget is fully spent or all firms in the observed subsets have been visited.
    Although any household can potentially visit any firm in the market, it can interact with only a fraction $F_h \in (0,1]$ of them in each time period. This friction is introduced to capture the presence of search costs. If $F_h = 1$,  then households can access all firms and consequently are likely to allocate their entire consumption budget to the lowest priced sellers, leaving the higher priced firms unable to sell their output. Instead, for lower values of $F_h$, where households can visit only a limited subset of firms, buyers may fail to exhaust their budgets, even though higher-priced sellers might experience higher sales than in the full-access scenario. This dynamic reveals a trade-off between demand rationing and unsold output, contingent upon the value of $F_h$.
	
    In the credit market, each borrower is initially matched with a lender based on a predetermined credit fitness value.
    Borrowers are sorted in ascending order by probability of default so that riskier firms are the first to be rationed when the available credit supply falls short of aggregate demand. This rule reflects the principle that, under credit supply constraints, banks aim to mitigate exposure by prioritizing lending to less risky counterparties.
    Firms enter the market one by one. A borrower firm $j$ can switch from lender $b$ to $z$, among those with positive credit supply, through a sampling mechanism with replacement with probability 
    $$
    p_{jb}^{\text{switch}} = \frac{1}{1 + \exp\left[-\kappa (\nu_b - \nu_z)\right]}\label{eq:placeholder}
    $$
    where weights $v_b$ and $v_z$ are determined from the credit fitness $x$ of each bank, $v = \frac{x}{\sum{x}}$. 
    If the credit demand of one firm cannot be satisfied by the lender to which it is attached, it is assigned to a random new lender. Each borrower can make $n^{cred}$ attempts to borrow.

    The interbank network is fully connected, meaning that any bank can visit any other. 
    The reservation rate for a borrower is the same among all banks.
    Borrowers enter the market in random order and bid a rate to a random seller. If the bid rate is greater than the reservation rate, borrowers can borrow the desired amount and adjust the bid rate downward, otherwise they revise the bid rate upward and try again with the next lender. All borrowers are allowed to make $n^{ibtent}$ offers in each session of the interbank market. Further details are described in Section \ref{sec:banks}.

    \subsubsection{Networks topology}
    \label{NETTOP}

    The financial architecture of the system is represented by a set of interconnected networks. 
    We model the deposit and shareholder networks, firm-bank credit network, and the interbank network. 

    The first two networks are static, i.e. links are predetermined at the model initialization stage and do not change, while the firm-bank credit network and the interbank network are dynamic, i.e. links change throughout the simulation.
    
    \begin{enumerate}
    \item Deposit networks specify the allocation of deposits by households and firms across banks. They are constructed through the following steps. For each agent, we draw the number of bank accounts from a Poisson distribution with parameter $\mu= 2$. We set the minimum number of links to one to ensure that everyone is connected to a bank. Next, households or firms are matched one by one to a set of randomly selected banks. Deposits are equally divided over links. Summary statistics for the deposit networks are reported in Table \ref{tab.deposit_ntw}.
    \begin{table}[h]
\centering
\scriptsize
\begin{tabular}{lccccc}
\toprule
  & mean & std & median & min & max \\
\midrule
firms links to banks & 2.1668 & 1.2495 & 2.0000 & 1.0000 & 9.0000 \\
households links to banks & 2.1172 & 1.2599 & 2.0000 & 1.0000 & 8.0000 \\
banks links to firms & 108.3400 & 8.0982 & 109.0000 & 90.0000 & 132.0000 \\
banks links to households & 531.9500 & 21.8120 & 535.0000 & 478.0000 & 598.0000 \\
\bottomrule
\end{tabular}
\caption{Summary statistics of the depositors' network. Statistics are computed for 10 independent random realizations of the networks.}
\label{tab.deposit_ntw}
\end{table}
    \item The shareholders’ network identifies the shareholders of firms and banks. Shareholders earn dividends from their affiliated firms and banks and bear bail-in responsibilities in case of bankruptcy. To match households and firms, we follow a similar mechanism as in the depositors’ network, but allow only half of the households to be shareholders, while the other half have zero links because they do not participate in the stock market.
    For banks, we create a network where the number of shareholders is proportional to a credit fitness parameters. This reflects the idea that banks with the best lending opportunities, measured by the credit fitness, are those that are more likely to grow in size and, therefore, have the largest number of shareholders. Summary statistics for the shareholders' networks are reported in Table \ref{tab.share_ntw}.
    \begin{table}[h]
\centering
\scriptsize
\begin{tabular}{lccccc}
\toprule
  & mean & std & median & min & max \\
\midrule
households links to firms & 1.0583 & 1.3736 & 0.5000 & 0.0000 & 8.0000 \\
households links to banks & 1.0683 & 1.3817 & 1.0000 & 0.0000 & 9.0000 \\
firms links to households & 5.2916 & 2.2500 & 5.0000 & 1.0000 & 14.0000 \\
banks links to households & 267.0800 & 302.3296 & 140.5000 & 53.0000 & 1144.0000 \\
\bottomrule
\end{tabular}
\caption{Summary statistics of the shareholders' networks. Statistics are computed on 10 independent random realizations of the networks.}
\label{tab.share_ntw}
\end{table}
    \item The firm-bank credit network captures the credit relationships between firms and banks. The network dynamically develops when banks and firms are matched in the credit market via a preferential attachment mechanism. This mechanism operates on the basis of a constant fitness score which determines attachment probabilities, and is randomly assigned to the banks at the initialization of the model, and remains fixed throughout the simulation. Credit fitness is generated from a power law distribution with an exponential cut-off, with probability density function and parameters $\gamma=3$ and $\lambda = 0.01$:
    $$
    p(x) = x^{-\gamma}\exp{(-\lambda x)}.
    $$
    As a result, the firm–bank network exhibits disassortative mixing, whereby a few highly connected nodes (banks) tend to link with many nodes (firms) that have low connectivity.

    \item Banks trade funds within a fully-connected  interbank network. Although the network arises endogenously through the interbank matching mechanism, it is strongly affected by the structure of the credit market. Banks are heterogeneous in their credit fitness parameter, which affects their lending opportunities to firms. As a result, the demand for interbank funds is proportional to the amount of credit lent to the real economy. This determines a topology where a few high-fitness banks demand interbank liquidity, while the remaining banks acts mostly as suppliers. 
    \end{enumerate}
    \end{subappendices}

\end{document}